\RequirePackage{lineno}
\newif\ifpreprint%
\ifpreprint%
\documentclass[10pt,preprint,english,aps,floatfix,amssymb,superscriptaddress,a4paper]{revtex4-2}
\else%
\documentclass[10pt,twocolumn,pra,english,floatfix,aps,amssymb,superscriptaddress,a4paper]{revtex4-2}
\fi%
\usepackage{ltxcmds}
\usepackage{amssymb, amsmath, mathtools}
\usepackage{dsfont}

\usepackage[english]{babel}
\usepackage[utf8]{inputenc}
\usepackage{blindtext}
\usepackage{graphicx}
\usepackage{lipsum}
\usepackage{siunitx} % units
\usepackage{dutchcal}

\usepackage{physics}
\usepackage{etoolbox}
\usepackage{xcolor}
\usepackage[normalem]{ulem}
\usepackage{pgfplots}
\renewcommand{\vec}[1]{\boldsymbol{#1}}

\newcommand{\ssm}{\rm \scriptscriptstyle}

\renewcommand{\phi}{\varphi}

\newcommand{\suppl}{Supplemental Material Sec.}
\newcommand{\suppls}{Supplemental Material Secs.}

\newcommand{\beginsupplement}{%
  \setcounter{section}{0}%
  \setcounter{subsection}{0}%
  \setcounter{equation}{0}%
  \setcounter{figure}{0}%
  \setcounter{table}{0}%
  \setcounter{page}{1}%
  \renewcommand{\thesection}{S\Roman{section}}%
  \renewcommand{\theequation}{S\arabic{equation}}%
  \renewcommand{\thefigure}{S\arabic{figure}}%
  \renewcommand{\thetable}{S\arabic{table}}%
}
\usepackage{hyperref}
\makeatletter
\let\origaddcontentsline\addcontentsline
\newcommand{\DisableTOC}{\renewcommand{\addcontentsline}[3]{}}
\newcommand{\EnableTOC}{\let\addcontentsline\origaddcontentsline}
\makeatother

\pgfplotsset{compat=1.18}
\hypersetup{colorlinks = true, linkcolor=magenta,citecolor=blue, urlcolor=magenta, bookmarksnumbered =  true}

\begin{document}
\DisableTOC

%TC:ignore

\title{Observation of ring states in a delicate topological insulator}

\author{C. Tornow}
\email{ctornow@phys.ethz.ch}
\affiliation{
Institute for Theoretical Physics, ETH Zurich, 8093 Z\"urich, Switzerland
}
\author{J. Rupprecht}
\affiliation{
Institute for Theoretical Physics, ETH Zurich, 8093 Z\"urich, Switzerland
}
\author{P. Engeler}
\affiliation{
Institute for Theoretical Physics, ETH Zurich, 8093 Z\"urich, Switzerland
}
\author{U. Drechsler}
\affiliation{
IBM Research - Europe, Zurich, R\"uschlikon, Switzerland
}
\author{K.~E. Huhtinen}
\affiliation{
Institute for Theoretical Physics, ETH Zurich, 8093 Z\"urich, Switzerland
}
\author{C. Devescovi}
\affiliation{
Institute for Theoretical Physics, ETH Zurich, 8093 Z\"urich, Switzerland
}
\author{S.~D. Huber}
% \email{sebastian.huber@phys.ethz.ch}
\affiliation{
Institute for Theoretical Physics, ETH Zurich, 8093 Z\"urich, Switzerland
}

\begin{abstract}
Topological insulators are typically characterized by particularly stable properties, such as global invariants, and can be identified by probing their robust surface states.
A recently discovered novel form of band topology, delicate topology, challenges this paradigm:
% Delicate topology is a novel, unusually subtle form of band topology: 
its defining property, multicellularity, can be removed by introducing a coupling to local orbitals anywhere in the spectrum, even far above the relevant band gap.
This makes it hard to diagnose delicate topology with conventional probes that access only low-energy degrees of freedom.
Here, we introduce strong local impurities as a spectroscopic probe of a delicate topological insulator which we realize in a phononic metamaterial.
By tuning the impurity strength and performing orbital-resolved readout, we observe recently proposed indicators of topology: \emph{ring states}, in-gap bound states whose frequencies remain pinned in the strong-impurity limit while their real-space profiles form a pronounced ring around the impurity site.
We find that these ring states persist even when the multicellularity in our system is removed by a weakly hybridizing additional orbital.
Our results establish impurity-induced ring states as probes of complex multiband physics, including delicate topological phases.
\end{abstract}

\maketitle

\ifpreprint%
	\linenumbers%
\fi%

%TC:endignore

%
\emph{Introduction---}Topological bands are defined by the global properties of Bloch wave functions and are typically characterized by bulk invariants or robust boundary phenomena~\cite{Hasan10, Qi2011}. These properties, in turn, can be observed in transport measurements~\cite{Konig07} or scanning probes~\cite{Roushan2009, Reis2017, Collins2018} that excel in carefully addressing the low-energy physics well below the scale of the band gap. These techniques are indeed optimally tailored to most conventional topological phenomena, such as the quantum Hall effect \cite{Klitzing80} or the quantum spin Hall effect \cite{Konig07, Bernevig06a}. 

However, many of the exciting recent developments in topological band theory draw essentially from high-energy aspects of the materials under investigation: 
Topological quantum chemistry \cite{Bradlyn17}, as a case in point, formulates band topology in terms of localized atomic orbitals. Their natural energy scales are not related to the intricate interference effects around the Fermi energy, which are on the order of a few millielectronvolts, but to the typical separation of atomic orbitals on the order of electronvolts. An important challenge in understanding such modern topological materials is, therefore, to find a reliable probe that can span these vastly different energy scales without destroying the sought-after low-energy physics.

Local impurities in the ultra-strong limit can serve exactly this purpose \cite{Queiroz2024}. Their local nature minimizes the effect on the low energy physics and ensures that the global topology cannot be ruined. 
The strength of the impurity, on the other hand, controls up to which energy scale the system is probed.
Here, we show how one can utilize the study of impurity-induced {\em ring states}~\cite{Queiroz2024} to investigate a novel, particularly subtle form of band topology: A delicate topological insulator \cite{Nelson2021, Nelson2022, Chen2024}.

What is a {\em delicate} topological insulator and why does it call for a novel probe? In a {\em stable} topological phase like the quantum Hall state \cite{Klitzing80}, the topology survives even if one introduces a coupling to additional atomic orbitals, e.g., when a two-dimensional topological material is brought into proximity to an additional layer. A slightly weaker form of topology is realized in a {\em fragile} topological insulator~\cite{Po2018, Song2020, Bouhon2019, Peri20}, where the phase can lose its topological character when additional orbitals appear below the Fermi energy. Delicate topology, finally, is even more subtle: Its defining property, multicellularity  \cite{Nelson2021}, i.e, the inability to describe the delicate bands with a set of orbitals residing in a single unit-cell, can disappear when orbitals are added anywhere in the Hilbert space.
% ; cf. Fig.~\ref{fig:Fig0}a\&b. 
In particular an additional orbital far {\em above} the band gap can remove the topology.  As a result of the relevance of the full Hilbert space, a reliable characterization of a delicate topological insulator cannot be based solely on the properties of the filled bands.
It requires a probe that covers the system's full Hilbert space.

In this work, we realize a delicate topological insulator in a two-dimensional silicon-based phononic metamaterial. To probe and characterize the phononic delicate topological insulator, we introduce local impurities of different strengths and record the impurity-induced response using orbital-resolved spectroscopy and real-space reconstruction of the corresponding modes.
Our main result is the observation of impurity-induced in-gap modes with the defining signatures of ring states: 
even for impurity strengths far beyond the band width the ring state frequencies remain pinned inside the topological gap, while their spatial profiles form the characteristic ring around the defect site.
% as the impurity strength is increased, its frequency remains pinned in the middle of the topological gap, while its spatial profile forms the characteristic ring around the defect site.
The ultra-strong limit of our impurities further exposes the full involved Hilbert space and therefore allows a complete characterization beyond the low-energy physics. 
This in turn brings forward unexpected insights into the complex multiband physics inevitably present in experimental systems.

\begin{figure}
    \centering
    \includegraphics[width=\columnwidth]{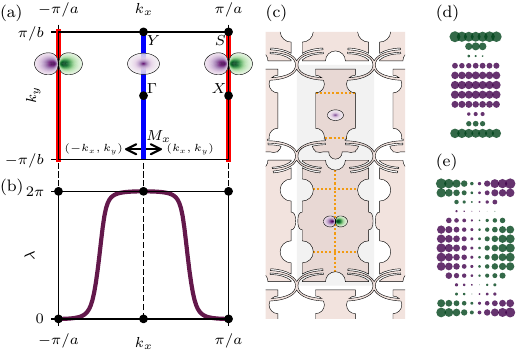}
     \caption{{Delicate topological insulator in a phononic metamaterial.\,} (a) Brillouin zone with high-symmetry points indicated. The lines $\overline{\Gamma Y}$ and $\overline{XS}$ are invariant under the mirror symmetry $M_x$. We consider a two-band model induced by a mirror even ($s$) and a mirror odd ($p$) orbital. In the lowest band the orbital character changes from the center ($s$, blue) to the edge of the Brillouin zone ($p$, red). (b) The Wilson loop phase $\lambda$, or equivalently, the center of the $y$-localized Wannier function of the lowest band as a function of $k_x$. For a two-band model with $M_x$-symmetry, the value of $\lambda=2\pi n$, $n\in \mathbb Z$ is quantized at the high-symmetry lines $k_x=0$ and $k_x=\pi/a$. Delicate topology (multicellularity) is realized when $\lambda$ is winding by $2\pi$ over half the Brillouin zone as shown. (c) A silicon (shaded in red) phononic metamaterial where each of the two plates in the shown unit cell (light gray) hosts one relevant mode; the nodes of the out-of-plane modes are indicated by the dashed orange lines. The thin arms lead to a coupling of these modes into a set of Bloch bands. (d), (e) Measured local modes on the $s$- and $p$-plate, respectively. The size of the circles indicate the squared local amplitude, the color denotes the sign of the out of plane response.}
     \label{fig:Fig0}
\end{figure}

\emph{Delicate topology in a phononic metamaterial---}We study a two-dimensional delicate topological insulator stabilized by the mirror symmetry $M_x: (x,y)\to (-x,y)$ and with a unit cell with lattice constants $(a,b)$. 
The delicate topology is induced by a band inversion between a mirror-even ($s$) and a mirror-odd ($p$) orbital. 
Their displacements are described by
 ${\vec z({\vec k})}=(z_s({\vec k}),z_p({\vec k}))$ in the form  $\ddot {\vec z}({\vec k}) = -\mathcal D(\vec k) {\vec z}({\vec k})$, where we write the dynamical matrix $\mathcal D(\vec k)$ as
\begin{equation}
\mathcal D(\vec k)=
\begin{bmatrix}
\mathcal s(\vec k) & \mathcal f_{sp}(\vec k)\\
\mathcal f_{sp}^*(\vec k) & \mathcal p(\vec k)
\end{bmatrix}.
\label{eq:dynamical_matrix_main}
\end{equation}
Here, $\mathcal s(\vec k)$ and $\mathcal p(\vec k)$ denote the uncoupled dispersions of the $s$ and $p$ orbitals, respectively, and $\mathcal f_{sp}(\vec k)$ denotes the inter-orbital coupling.
Mirror symmetry enforces that the inter-orbital coupling vanishes along the mirror-invariant (high-symmetry) lines $k_x=0$ and  $k_x=\pi/a$, i.e., $\mathcal f_{sp}(0,k_y)=\mathcal f_{sp}(\pi/a,k_y)=0$.
Band inversion is realized by the additional constraint that $\mathcal s(0,k_y)<\mathcal p(0,k_y)$ and $\mathcal s(\pi/a,k_y)>\mathcal p(\pi/a,k_y)$.
Thus, along $k_x = 0$ the lower band is purely $s$-like, while along $k_x = \pi/a$ it is $p$-like [Fig.~\ref{fig:Fig0}(a)]. 
This constraint quantizes the Wilson loop phase $\lambda$~\cite{Wilczek1984, Zak1982}, or equivalently, the center of the $y$-localized Wannier function of each band~\cite{Coh2009}, at the high-symmetry lines to values $\lambda(k_x=0)=2\pi l$ and $\lambda(k_x=\pi/a)=2\pi m$, with $l,m\in \mathbb Z$ (\suppl~\ref{sec:wilson}). 
In order to be in the delicate topological phase, the coupling between the two orbitals $\mathcal f_{sp}(\vec k)$ needs to open a gap throughout the whole Brillouin zone and induce a winding of the Wilson loop phase with $|\lambda(k_x=\pi/a) - \lambda(k_x=0)| = 2\pi|m-l|$, where $l \neq m$~\cite{Nelson2021, Nelson2022, Chen2024, Cheng2025}. 

A possible realization with $|m-l|=1$, cf. Fig.~\ref{fig:Fig0}b, is given by coupling the $s$ and $p$ orbital in a Su-Schrieffer-Heeger~\cite{Su79} (SSH) manner along $y$, as described by:
\begin{equation}
\mathcal f_{sp}(\vec k) = 2{i}\sin(k_x a)\left[t_{sp}^{10}+t_{sp}^{11\uparrow} e^{{\rm i}k_y b}\right].
\end{equation}
Here, $t_{sp}^{10}$ and $t_{sp}^{11\uparrow}$ parametrize effective inter-orbital couplings between neighboring unit cells (\suppl~\ref{sec:full_tb_model}).

In Fig.~\ref{fig:Fig0}c, we present a mechanical structure that realizes such a setup where the two depicted plates host an $s$- and a $p$-mode, respectively. The elliptic arms induce in leading order the sought after couplings (\suppl~\ref{sec:sample_design}).
% ~\ref{sec:effective_model}).
Experimentally, we resolve the local mode shapes on each plate [Fig.~\ref{fig:Fig0}(d), (e)], which later enables orbital-resolved band structure measurements and impurity spectroscopy.
We call the plate hosting the $s$-mode and the $p$-mode, ``$s$-plate" and ``$p$-plate" in the following.

\begin{figure}
    \centering
    \includegraphics[width=\columnwidth]{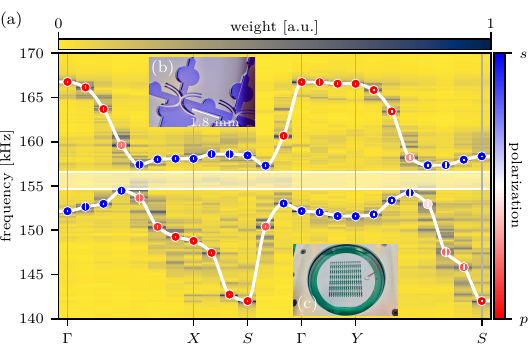}
    \caption{{Dispersion along high-symmetry lines.} (a) Measured response of the out-of-plane excitation as a function of wave vector and frequency along high-symmetry lines in the Brillouin zone. The density plot in the background depicts the raw response data. The dots show the location of the peaks, the color of each dot indicates the experimentally determined $s$/$p$ polarization while the error bars denote the peak widths extracted from Lorentzian fits (\suppl~\ref{sec:orbital_resolved_bulk_measurements}). Along the lines $\overline{XS}$ and $\overline{\Gamma Y}$ the modes are fully polarized as required by mirror symmetry. 
    White shaded regions indicate the experimentally verified range of the band gap (\suppl~\ref{sec:orbital_resolved_bulk_measurements}).
    (b) Micrograph of the connecting arms fabricated by through-wafer deep reactive ion etching. (c) Photo of a full sample on a 4 inch wafer mounted inside the vacuum chamber and fitted with an exciter piezoelectric actuator. 
    % No signifcant response is hidden by the insets.
    }
    \label{fig:Fig1}
\end{figure}

\emph{Orbital-resolved bulk response---}We realize the mechanical structure with $15 \times 6$ unit cells in a silicon wafer using through-wafer deep reactive ion etching, see Fig.~\ref{fig:Fig1}(b), (c).
The detailed fabrication process is presented in \suppl~\ref{sec:fabrication}.
We excite the sample with a piezoelectric actuator and perform real-time lock-in measurements of the out-of-plane motion of single plates using laser interferometry.
Orbital-resolved measurements are realized by detecting the amplitude and phase on different points on each of the plates and post-processing the information (\suppl~\ref{sec:orbital_resolved_measurements_normalization}).

By performing a spatial Fourier transform of the measured lattice response and tracking orbital-resolved resonances along a high-symmetry path through the Brillouin zone, we reconstruct the bulk dispersion (\suppl~\ref{sec:orbital_resolved_bulk_measurements}).
% ~\ref{sec:orbital_resolved_bulk_measurements}).
The measured band structure resolves two bands [Fig.~\ref{fig:Fig1}(a)] in the expected frequency range with a gap of about $\SI{1.9}{\kilo \hertz}$ (\suppls~\ref{sec:sample_design} and \ref{sec:orbital_resolved_bulk_measurements}).
The bands exchange their $s$/$p$ polarization across the Brillouin zone as expected.

Before we move on to the discussion of the experimental observation of ring states, we provide a short summary of the theoretical description of local impurities and induced states in multiband systems.

\emph{Ring states as probes of complex multiband physics---}In non-interacting systems, the single-impurity problem can be solved exactly. 
We consider such a system described by a multiband Hamiltonian or dynamical matrix $\mathcal D(\vec k)$ and an impurity with strength $U_\alpha$ that acts on a single local orbital $\alpha$. 
The impurity can be described by the potential $U_\alpha\ket{\alpha}\bra{\alpha}$, where $\ket{\alpha}$ denotes a local orbital.
We study the response of the system to such a local impurity in the strong-impurity limit $|U_\alpha| \rightarrow \infty$.
 
In general, and independent of the details of $\mathcal D(\vec k)$, one impurity-induced bound state is pulled out of the spectrum of $\mathcal D(\vec k)$ to very high or very low energies, depending on the sign of $U_\alpha$, as the impurity strength increases.
The energy $\epsilon_{\rm b}$ of this state scales with the strength of the impurity, i.e., $\epsilon_{\rm b} \sim U_\alpha$~\cite{Queiroz2024}. 
We call this state an {\em impurity state} since its wave function coincides with $\ket{\alpha}$ up to corrections of $1/U_\alpha$ and therefore has dominant weight on the impurity site~\cite{Queiroz2024}.

In contrast, an impurity-induced bound state inside a band gap, whose energy remains at a finite value even in the strong-impurity limit, is controlled by the nature of the unperturbed system described by $\mathcal D(\vec k)$ rather than the microscopic details of the impurity potential.
Fundamentally, such a state exists only if the impurity-projected Green's function $g_\alpha(\epsilon)$, a property of the unperturbed system, crosses zero at an energy $\epsilon^{*}_{\rm b}$ inside the bulk gap (\suppl~\ref{sec:greens_function_numerics}).
% The associated mode is orthogonal to the bare impurity eigenstate. 
It avoids the impurity site by developing a characteristic ring-shaped spatial profile with enhanced weight on the sites surrounding the impurity site, hence the name {\em ring states}~\cite{Queiroz2024}.

Ring states have initially been proposed as indicators of topology~\cite{Queiroz2024}.
In many band insulators, topology is accompanied by a band inversion within an effective two-band subspace (e.g., spanned by two orbitals $\alpha$ and $\beta$).
This implies that these bands cannot be individually represented by symmetry-preserving exponentially localized orbitals: both bands necessarily have non-vanishing overlap with both orbitals $\alpha$ and $\beta$.
If an impurity acts on one of the participating orbitals, for instance $\alpha$, the corresponding impurity-projected Green's function $g_\alpha$ is forced to cross zero in the topological band gap~\cite{Queiroz2024, Misawa2022}.
As a consequence, the existence of impurity-induced ring states is guaranteed.
Such Green's function zeros and the associated ring states are protected against symmetry-preserving adiabatic deformations~\cite{Queiroz2024}.

Because strong impurities effectively probe the full Hilbert space, ring states can further serve as powerful probes of systems governed by complex multiband physics beyond two-band topology~\cite{Diop2020, Misawa2022, Pangburn2025}.
In the following, we show experimentally how ring states not only provide insight into the underpinning of delicate topology but also its lifting by giving access to the system's full Hilbert space.

\begin{figure*}[!htbp]
    \centering
\includegraphics[width=\textwidth]{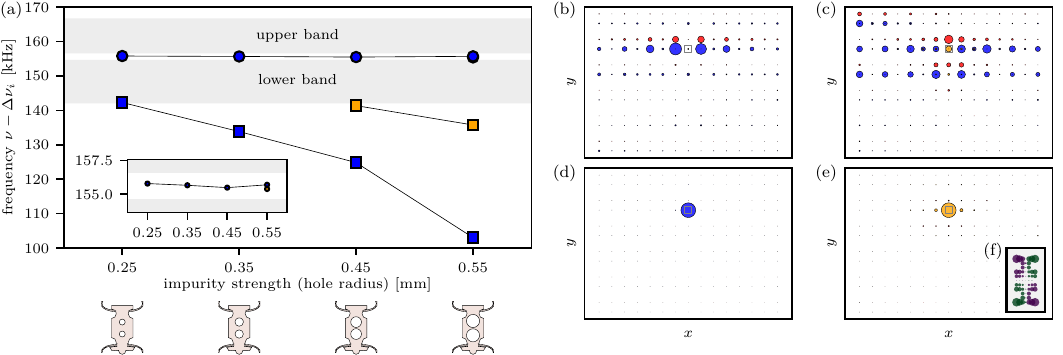}
    \caption{{Experimental observation of impurity-induced ring states.} 
    (a) Measured eigenfrequencies of impurity-induced states as a function of impurity strength.
    Impurities are realized by two circular holes of increasing radius $r\in \{0.25, 0.35, 0.45, 0.55 \}\,{\rm mm}$ in single $s$-plates.
    Each resonance frequency is determined by a Lorentzian fit to a resonance peak.
    All frequencies are plotted with respect to the bulk sample spectrum (cf. Fig.~\ref{fig:Fig1}) by compensating the overall sample-dependent frequency shift $\Delta\nu_i$, where the index $i$ denotes an impurity sample (\suppl~\ref{sec:app_bandstructure_mapping}).
    Error bars correspond to the uncertainties of $\Delta\nu_i$ and are smaller than the marker sizes.
    Shaded regions indicate the frequency ranges of the lower and upper bulk bands (\suppl~\ref{sec:orbital_resolved_bulk_measurements}).
    Blue symbols denote resonance frequencies corresponding to $s$ modes: circles mark the in-gap ring states in the band gap between lower and upper band, squares the impurity states which leave the lower band and move to lower frequencies with increasing hole radius.
    The orange circle shows an in-gap $\tilde{p}$ mode resonance and orange squares display the impurity states associated with the $\tilde{p}$ orbital.
    % The error bars correspond to the standard deviation of the applied sample-dependent frequency shift (\suppl~\ref{sec:app_bandstructure_mapping}). 
    (b), (c) Spatial profiles of the $\SI{0.55}{\milli \meter}$ $s$ impurity-induced ring state and the $\SI{0.55}{\milli \meter}$ $\tilde{p}$ impurity-induced in-gap state. 
    (d), (e) Spatial profiles of the $\SI{0.55}{\milli \meter}$ $s$ impurity state and the $\SI{0.55}{\milli \meter}$ $\tilde{p}$ impurity state. 
    The circle areas in (b)--(e) are directly proportional to the measured and normalized squared amplitude of each mode (\suppls~\ref{sec:signal_analysis} and \ref{sec:orbital_resolved_measurements_normalization}).
    Blue, red and orange circles correspond to $s$, $p$ and $\tilde{p}$ modes.
    The spatial profiles of all $s$ impurity-induced ring and impurity states are shown in Fig.~\ref{fig:FigSI5} (\suppl~\ref{sec:impurity_spectroscopy}).
    (f) Experimentally determined local $\tilde p$ mode on an $s$ plate. Circle size and color correspond to the squared local amplitude and sign of the out-of-plane response, respectively.
    }
    \label{fig:Fig2}
\end{figure*}

\emph{Observation of impurity-induced ring states---}To study the response of the delicate topological insulator to local impurities we fabricate impurity samples, each hosting several impurities of different strengths.
The impurities are realized by circular holes in selected $s$ plates of the lattice. By increasing the hole radius $r$ we tune the local stiffness and mass of the plate and thereby shift the corresponding on-site $s$ mode frequency to lower values, effectively realizing an attractive on-site potential $U_s \ket{s} \bra{s}$ for the $s$ orbital. The impurity $s$ plates are schematically shown at the bottom of Fig.~\ref{fig:Fig2}(a).

We show the measured impurity-induced state frequencies relative to the bulk bands in Fig.~\ref{fig:Fig2}(a) (see also \suppl~\ref{sec:impurity_spectroscopy}). As expected, each impurity produces a low-frequency impurity state, which is pulled out of the lower band and rapidly shifts downward in frequency with increasing impurity strength (by $\SI{39.3 \pm 0.2}{\kilo \hertz}$ across the full range).
We reconstruct the spatial profile of this mode in Fig.~\ref{fig:Fig2}(d) and confirm that this state is localized on the impurity site.

Inside the bulk band gap we observe a single $s$-mode resonance per impurity. 
Their frequencies vary only weakly, by at most $\SI{0.30 \pm 0.23}{\kilo \hertz}$ across the full range of impurity strengths, as shown in the inset of Fig.~\ref{fig:Fig2}(a).
In Fig.~\ref{fig:Fig2}(b) we present the detailed measurement of the mode profile of such an in-gap state. As predicted for ring states, the excitation is strongly suppressed at the impurity plate and has an enhanced weight on surrounding sites.

We confirm our observation of ring states by evaluating the impurity-projected Green's function $g_s(\epsilon)$ for our two-band model [Eq.~(\ref{eq:dynamical_matrix_main}); \suppls~\ref{sec:full_tb_model} and \ref{sec:greens_function_numerics}].
Indeed, we find that $g_s(\epsilon)$ crosses zero within the band gap, which implies the existence of ring states pinned to the zero-crossing energy $\epsilon_{\rm b}^{*}$ in the strong impurity limit.

\begin{figure}[!htbp]
    \centering
\includegraphics[width=\columnwidth]{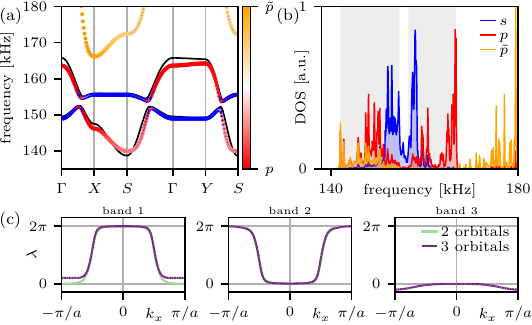}
    \caption{{Three band model.}
    (a) Energy eigenvalues along high-symmetry lines in the Brillouin zone of a three-band tight-binding Hamiltonian with parameters extracted from finite-element simulations (\suppl~\ref{sec:full_tb_model}).
    The size of the colored dots indicates the squared overlap of the corresponding eigenmode with the $s$ orbital and the mirror-odd orbital subspace ${\rm span}\{p, \tilde p\}$, shown in blue and by the red-orange color map, respectively.
    Within the mirror-odd sector the color map encodes the overlap with each of the mirror-odd modes, where red corresponds to maximal overlap with the $p$ orbital and orange to maximal overlap with the $\tilde p$ orbital.
    The black lines indicate the solution obtained from a finite-element simulation of the structure with periodic boundary conditions (\suppl~\ref{sec:effective_model_extraction}).
    (b) Experimentally determined orbital-resolved density of states for the three orbitals.
    Grey shaded regions indicate the range of the two lower bands [cf. Fig.~\ref{fig:Fig1}; \suppl~\ref{sec:orbital_resolved_bulk_measurements}].
    (c) Wilson loop phase $\lambda(k_x)$ (\suppl~\ref{sec:wilson}) for the Bloch eigenstates versus $k_x$ for the three bands in (a), comparing the model with two orbitals (solid green) to the model with three orbitals per unit cell (dashed purple).}
    \label{fig:Fig3}
\end{figure}

\emph{Trivializing delicate topology---}Besides the $s$-like ring and impurity states, the measurements also reveal additional resonances (labeled as orange markers in Fig.~\ref{fig:Fig2}(a)) that predominantly occupy a second, mirror-odd mode on the $s$ plate, which we denote $\tilde{p}$ [Fig.~\ref{fig:Fig2}(f)].
Impurities on the $s$-plates perturb both the $s$ and $\tilde p$ modes. %, i.e., the local impurity is rank-2.
Experimentally, we observe two $\tilde p$-dominated impurity-induced modes. One is pulled out of the lowest band with increasing impurity strength [Fig.~\ref{fig:Fig2}(a), (e)] and one state appears in the gap for the highest impurity strength [inset of Fig.~\ref{fig:Fig2}(a), (c)]. These modes are not captured by a minimal two-band description and signal the presence of a third, higher-lying band which weakly hybridizes with the $s$/$p$ subspace.

To capture this multiband physics, we construct a three-band tight-binding model including the $s$, $p$ and $\tilde p$ orbitals, with parameters extracted from finite-element simulations (\suppl~\ref{sec:full_tb_model}).
The two lowest bands retain the mirror-enforced band inversion between mirror-even and mirror-odd modes that underlies the delicate topological phase in the effective two-band description, whereas the third band is predominantly $\tilde p$-like [Fig.~\ref{fig:Fig3}(a)].
This behavior is experimentally verified by measuring the orbital-resolved density of states for the bulk sample [Fig.~\ref{fig:Fig3}(b)].
We find that the solution of the three-band tight-binding model is in good agreement with the band structure extracted from finite-element simulations of the structure with periodic boundary conditions (black lines in Fig.~\ref{fig:Fig3}(a); \suppl~\ref{sec:effective_model_extraction}).

From the perspective of band topology, the third orbital has a decisive effect: 
On the mirror-invariant lines, mixing within the mirror-odd sector ${\rm span}\{p, \tilde p\}$ is allowed.
As a consequence, the Wilson-loop phase quantization of the two lower bands, i.e, multicellularity, is lost, and hence the delicate topology is trivialized [Fig.~\ref{fig:Fig3}(c); \suppl~\ref{sec:wilson}].

The observation of ring states, even in the absence of delicate topology, and the appearance of an additional $\tilde p$ in-gap state and impurity resonances motivates an investigation of the complex multiband physics involved.
The first question we aim to answer is: Is the third band purely accidental, i.e., can it be removed in a symmetry-preserving adiabatic transformation of the three-band model?
Indeed, we show that the $\tilde p$ band can be smoothly decoupled without closing the gap of the lower two bands, hence retrieving an effective $s$/$p$ two-band delicate topological insulator with a quantized Wilson loop phase (\suppl~\ref{sec:adiabatic_transformations}). 

A second question arises: Are the $p$ and $\tilde p$ orbitals in our three-band system exchangeable, i.e, can we find a gapped interpolation that connects the effective $s$/$p$ system to an effective $s$/$\tilde p$ two-band delicate topological model?
Indeed, we show that such an interpolation exists (\suppl~\ref{sec:hot_swap}).
The $s$ ring states persist even in this limit, as we confirm by evaluating the impurity-projected Green's function $g(\epsilon)$ in the impurity subspace ${\rm span}\{ s, \tilde p\}$ (\suppl~\ref{sec:hot_swap}).
Fundamentally, these ring states probe the band inversion between the mirror-even $s$ orbital and mirror-odd subspace ${\rm span}\{p, \tilde p\}$, which remains intact.
In contrast, in our physical realization, the additional $\tilde p$ in-gap state is not associated with an in-gap zero of $g(\epsilon)$ and therefore does not correspond to a pinned ring state (\suppl~\ref{sec:greens_function_numerics}).

\emph{Discussion---}We have experimentally observed and characterized ring states in a two-dimensional silicon phononic metamaterial implementing a mirror-symmetry stabilized delicate topological insulator.
Our experimental results sharpen the interpretation of impurity spectroscopy as a probe of topology.
Consistent with previous theoretical studies~\cite{Diop2020, Misawa2022, Pangburn2025} and extending related work on ordinary defects and bound states in topological insulators~\cite{Liu09,Shan2011,Naselli2022,Michel2024,Mains2025}, we find that ring states do not uniquely probe a topological invariant: instead they diagnose the existence and robustness of impurity-projected Green's function zeros which can be enforced in topologically non-trivial systems but more precisely track band inversion and mixing in multiband systems.
By systematically increasing the impurity strength, our experiment provides a clean metamaterial realization of this physics that has not been accessed so far in electronic materials.

Strong-impurity spectroscopy offers a route to disentangling complex spectra and identifying the relevant degrees of freedom even in fragile or delicate systems when conventional low-energy probes may become ambiguous.
This potentially opens interesting experimental avenues also beyond phononic metamaterials such as in cold atoms, photonics or conventional electronic condensed matter systems.

\emph{Acknowledgments---}The authors thank R. Queiroz and T. Prijon for insightful discussions and M. Friedrichs and L. Schneider for their technical contributions to the experimental setup.

\bibliographystyle{phd-url}
\bibliography{ref}

@article{Nelson2021,
	author = {Nelson, A. and Neupert, T. and Bzdu{\v s}ek, T. and Alexandradinata, A.},
	doi = {10.1103/PhysRevLett.126.216404},
	journal = {Phys. Rev. Lett.},
	pages = {216404},
	title = {Multicellularity of Delicate Topological Insulators},
	url = {https://doi.org/10.1103/PhysRevLett.126.216404},
	volume = {126},
	year = {2021}}

@article{Chen2024,
	author = {Chen, Y.-C. and Lin, Y.-P. and Kao, Y.-J.},
	doi = {10.1038/s42005-023-01502-8},
	journal = {Commun. Phys.},
	pages = {32},
	title = {Chern dartboard insulator: sub-Brillouin zone topology and skyrmion multipoles},
	url = {https://doi.org/10.1038/s42005-023-01502-8},
	volume = {7},
	year = {2024}}

@article{Nelson2022,
	author = {Nelson, A. and Neupert, T. and Alexandradinata, A. and Bzdu{\v s}ek, T.},
	journal = {Phys. Rev. B},
	month = {PhysRevB.106.075124.pdf},
	pages = {075124},
	title = {Delicate topology protected by rotation symmetry: Crystalline Hopf insulators and beyond},
	volume = {106},
	year = {2022},
    url = {https://link.aps.org/doi/10.1103/PhysRevB.106.075124}}

@article{Song2020,
	author = {Song, Z.-D. and Elcoro, L. and Bernevig, B. A.},
	journal = {Science},
	pages = {794},
	title = {Twisted bulk-boundary correspondence of fragile topology},
	volume = {367},
	year = {2020},
    url = {https://www.science.org/doi/abs/10.1126/science.aaz7650}}

@article{Peri20,
	author = {Peri, V. and Song, Z. and Serra-Garcia, M. and Engeler, P. and Queiroz, R. and Huang, X. and Deng, W. and Liu, Z. and Bernevig, B. A. and Huber, S. D.},
	journal = {Science},
	pages = {797},
	title = {Experimental characterization of spectral flow between fragile bands},
	volume = {367},
	year = {2020},
    url = {https://www.science.org/doi/abs/10.1126/science.aaz7654}}

@article{Bradlyn17,
	author = {Bradlyn, B. and Elcoro, L. and Cano, J. and Vergniory, M.G. and Wang, Z. and Felser, C. and Aroyo, M.I. and Bernevig, B. A.},
	journal = {Nature},
	pages = {298},
	title = {Topological quantum chemistry},
	volume = {537},
	year = {2017},
    url = {https://doi.org/10.1038/nature23268}}

@article{Serra-Garcia18,
	author = {Serra-Garcia, M. and V. Peri and S{\"u}sstrunk, R. and Bilal, O. R. and Larsen, T. and Villanueva, L.G. and Huber, S. D.},
	journal = {Nature},
	pages = {342},
	title = {Observation of a phononic quadrupole insulator},
	volume = {555},
	year = {2018},
    url = {https://doi.org/10.1038/nature25156}}

@article{Matlack18,
	author = {Matlack, K. H. and Serra-Garcia, M. and Palermo, A. and Huber, S. D. and Daraio, C.},
	journal = {Nature Mat.},
	pages = {323},
	title = {Designing perturbative metamaterials from discrete models},
	volume = {17},
	year = {2018},
    url = {https://doi.org/10.1038/s41563-017-0003-3}}

@article{Su79,
	author = {Su, W. P. and Schrieffer, J. R. and Heeger, A. J.},
	journal = {Phys. Rev. Lett.},
	pages = {1698},
	title = {Solitons in Polyacetylene},
	volume = {42},
	year = {1979},
    url = {https://link.aps.org/doi/10.1103/PhysRevLett.42.1698}}

@article{Hasan10,
	author = {Hasan, M. Z. and Kane, C. L.},
	journal = {Rev. Mod. Phys.},
	pages = {3045},
	title = {Colloquium: Topological insulators},
	volume = {82},
	year = {2010},
    url = {https://doi.org/10.1103/RevModPhys.82.3045}}

@article{Klitzing80,
	author = {v. Klitzing, K. and Dorda, G. and Pepper, M.},
	journal = {Phys. Rev. Lett.},
	pages = {494},
	title = {New Method for High-Accuracy Determination of the Fine-Structure Constant Based on Quantized {Hall} Resistance},
	volume = {45},
	year = {1980},
    url = {https://link.aps.org/doi/10.1103/PhysRevLett.45.494}}

@article{Konig07,
	author = {K{\"o}nig, M. and Wiedmann, S. and Br{\"u}ne, C. and Roth, A. and Buhman, H. and Molenkamp, L. W. and Qi, X.-L. and Zhang, S.-C.},
	journal = {Science},
	pages = {766},
	title = {Quantum Spin Hall Insulator State in HgTe Quantum Wells},
	volume = {318},
	year = {2007},
    url = {https://www.science.org/doi/abs/10.1126/science.1148047}}

@article{Bernevig06a,
	author = {Bernevig, B. A. and Hughes, T. L. and Zhang, S.-C.},
	journal = {Science},
	pages = {1757},
	title = {Quantum Spin Hall Effect and Topological Phase Transition in HgTe Quantum Wells},
	volume = {314},
	year = {2006},
    url = {https://www.science.org/doi/abs/10.1126/science.1133734}}

@article{Queiroz2024,
	author = {Raquel Queiroz and Roni Ilan and Zhida Song and B. Andrei Bernevig and Ady Stern},
	journal = {arXiv:2406.03529},
	title = {Ring states in topological materials},
	year = {2024},
    url = {https://arxiv.org/abs/2406.03529}}

@article{Mains2025,
    author={Aiden J. Mains and Jia-Xin Zhong and Yun Jing and Bitan Roy},
    journal = {arXiv:2511.10646},
    title={Ordinary lattice defects as probes of topology},
    year={2025},
    url={https://arxiv.org/abs/2511.10646}
}

@article{Pangburn2025,
    title = {Impurity-induced Mott ring states and Mott zeros ring states in the Hubbard operator formalism},
    author = {Pangburn, Emile and Banerjee, Anurag and P\'epin, Catherine and Bena, Cristina},
    journal = {Phys. Rev. B},
    volume = {112},
    issue = {12},
    pages = {125157},
    numpages = {20},
    year = {2025},
    publisher = {American Physical Society},
    doi = {10.1103/lysb-n9zp},
    url ={https://doi.org/10.1103/lysb-n9zp}
}

@article{Cheng2025,
	author = {Cheng, Zheyu and Yue, Sijie and Long, Yang and Xie, Wentao and Yu, Zixuan and Teo, Hau Tian and Zhao, Y. X. and Xue, Haoran and Zhang, Baile},
	journal = {Nature Comm.},
	pages = {9669},
	title = {Observation of returning Thouless pumping},
	volume = {16},
	year = {2025},
    url = {https://doi.org/10.1038/s41467-025-64671-w}}

@article{Diop2020,
    title = {Impurity bound states as detectors of topological band structures revisited},
    author = {Diop, Seydou-Samba and Fritz, Lars and Vojta, Matthias and Rachel, Stephan},
    journal = {Phys. Rev. B},
    volume = {101},
    issue = {24},
    pages = {245132},
    numpages = {12},
    year = {2020},
    publisher = {American Physical Society},
    doi = {10.1103/PhysRevB.101.245132},
    url = {https://link.aps.org/doi/10.1103/PhysRevB.101.245132}
}

@article{Michel2024,
  title = {Bound states and local topological phase diagram of classical impurity spins coupled to a Chern insulator},
  author = {Michel, Simon and F\"unfhaus, Axel and Quade, Robin and Valent\'{\i}, Roser and Potthoff, Michael},
  journal = {Phys. Rev. B},
  volume = {109},
  issue = {15},
  pages = {155116},
  numpages = {14},
  year = {2024},
  publisher = {American Physical Society},
  doi = {10.1103/PhysRevB.109.155116},
  url = {https://link.aps.org/doi/10.1103/PhysRevB.109.155116}
}

@article{Reis2017,
	author = {F. Reis and G. Li and L. Dudy and M. Bauernfeind and S. Glass and W. Hanke and R. Thomale and J. Schäfer and R. Claessen},
	journal = {Science},
	pages = {287-290},
	title = {Bismuthene on a SiC substrate: A candidate for a high-temperature quantum spin Hall material},
	volume = {357},
	year = {2017},
    url = {https://www.science.org/doi/abs/10.1126/science.aai8142}}

@article{Roushan2009,
	author = {Roushan, Pedram and Seo, Jungpil and Parker, Colin V. and Hor, Y. S. and Hsieh, D. and Qian, Dong and Richardella, Anthony and Hasan, M. Z. and Cava, R. J. and Yazdani, Ali},
	journal = {Nature},
	pages = {1106--1109},
	title = {Topological surface states protected from backscattering by chiral spin texture},
	volume = {460},
	year = {2009},
    url = {https://doi.org/10.1038/nature08308}}

@article{Collins2018,
	author = {Collins, James L. and Tadich, Anton and Wu, Weikang and Gomes, Lidia C. and Rodrigues, Joao N. B. and Liu, Chang and Hellerstedt, Jack and Ryu, Hyejin and Tang, Shujie and Mo, Sung-Kwan and Adam, Shaffique and Yang, Shengyuan A. and Fuhrer, Michael S. and Edmonds, Mark T.},
	journal = {Nature},
	pages = {390--394},
	title = {Electric-field-tuned topological phase transition in ultrathin Na3Bi},
	volume = {564},
	year = {2018},
    url = {https://doi.org/10.1038/s41586-018-0788-5}}

@article{Qi2011,
  title = {Topological insulators and superconductors},
  author = {Qi, Xiao-Liang and Zhang, Shou-Cheng},
  journal = {Rev. Mod. Phys.},
  volume = {83},
  issue = {4},
  pages = {1057--1110},
  numpages = {0},
  year = {2011},
  month = {Oct},
  publisher = {American Physical Society},
  doi = {10.1103/RevModPhys.83.1057},
  url = {https://link.aps.org/doi/10.1103/RevModPhys.83.1057}
}

@article{Po2018,
  title = {Fragile Topology and Wannier Obstructions},
  author = {Po, Hoi Chun and Watanabe, Haruki and Vishwanath, Ashvin},
  journal = {Phys. Rev. Lett.},
  volume = {121},
  issue = {12},
  pages = {126402},
  numpages = {6},
  year = {2018},
  month = {Sep},
  publisher = {American Physical Society},
  doi = {10.1103/PhysRevLett.121.126402},
  url = {https://link.aps.org/doi/10.1103/PhysRevLett.121.126402}
}

@article{Bouhon2019,
  title = {Wilson loop approach to fragile topology of split elementary band representations and topological crystalline insulators with time-reversal symmetry},
  author = {Bouhon, Adrien and Black-Schaffer, Annica M. and Slager, Robert-Jan},
  journal = {Phys. Rev. B},
  volume = {100},
  issue = {19},
  pages = {195135},
  numpages = {25},
  year = {2019},
  month = {Nov},
  publisher = {American Physical Society},
  doi = {10.1103/PhysRevB.100.195135},
  url = {https://link.aps.org/doi/10.1103/PhysRevB.100.195135}
}

@article{Khalaf2021,
  title = {Boundary-obstructed topological phases},
  author = {Khalaf, Eslam and Benalcazar, Wladimir A. and Hughes, Taylor L. and Queiroz, Raquel},
  journal = {Phys. Rev. Res.},
  volume = {3},
  issue = {1},
  pages = {013239},
  numpages = {38},
  year = {2021},
  month = {Mar},
  publisher = {American Physical Society},
  doi = {10.1103/PhysRevResearch.3.013239},
  url = {https://link.aps.org/doi/10.1103/PhysRevResearch.3.013239}
}

@article{Shan2011,
  title = {Vacancy-induced bound states in topological insulators},
  author = {Shan, Wen-Yu and Lu, Jie and Lu, Hai-Zhou and Shen, Shun-Qing},
  journal = {Phys. Rev. B},
  volume = {84},
  issue = {3},
  pages = {035307},
  numpages = {6},
  year = {2011},
  month = {Jul},
  publisher = {American Physical Society},
  doi = {10.1103/PhysRevB.84.035307},
  url = {https://link.aps.org/doi/10.1103/PhysRevB.84.035307}
}

@article{Naselli2022,
  title = {Nontrivial gapless electronic states at the stacking faults of weak topological insulators},
  author = {Naselli, Gabriele and K\"onye, Viktor and Das, Sanjib Kumar and Angilella, G. G. N. and Isaeva, Anna and van den Brink, Jeroen and Fulga, Cosma},
  journal = {Phys. Rev. B},
  volume = {106},
  issue = {9},
  pages = {094105},
  numpages = {8},
  year = {2022},
  month = {Sep},
  publisher = {American Physical Society},
  doi = {10.1103/PhysRevB.106.094105},
  url = {https://link.aps.org/doi/10.1103/PhysRevB.106.094105}
}

@article{Liu09,
  title = {Classical spins in topological insulators},
  author = {Liu, Qin and Ma, Tianxing},
  journal = {Phys. Rev. B},
  volume = {80},
  issue = {11},
  pages = {115216},
  numpages = {5},
  year = {2009},
  month = {Sep},
  publisher = {American Physical Society},
  doi = {10.1103/PhysRevB.80.115216},
  url = {https://link.aps.org/doi/10.1103/PhysRevB.80.115216}
}

@article{Zak1982,
  title = {Band Center---A Conserved Quantity in Solids},
  author = {Zak, J.},
  journal = {Phys. Rev. Lett.},
  volume = {48},
  issue = {5},
  pages = {359--362},
  numpages = {0},
  year = {1982},
  month = {Feb},
  publisher = {American Physical Society},
  doi = {10.1103/PhysRevLett.48.359},
  url = {https://link.aps.org/doi/10.1103/PhysRevLett.48.359}
}

@article{Coh2009,
  title = {Electric Polarization in a Chern Insulator},
  author = {Coh, Sinisa and Vanderbilt, David},
  journal = {Phys. Rev. Lett.},
  volume = {102},
  issue = {10},
  pages = {107603},
  numpages = {4},
  year = {2009},
  month = {Mar},
  publisher = {American Physical Society},
  doi = {10.1103/PhysRevLett.102.107603},
  url = {https://link.aps.org/doi/10.1103/PhysRevLett.102.107603}
}

@article{Wilczek1984,
  title = {Appearance of Gauge Structure in Simple Dynamical Systems},
  author = {Wilczek, Frank and Zee, A.},
  journal = {Phys. Rev. Lett.},
  volume = {52},
  issue = {24},
  pages = {2111--2114},
  numpages = {0},
  year = {1984},
  month = {Jun},
  publisher = {American Physical Society},
  doi = {10.1103/PhysRevLett.52.2111},
  url = {https://link.aps.org/doi/10.1103/PhysRevLett.52.2111}
}

@article{Misawa2022,
  title = {Zeros of Green functions in topological insulators},
  author = {Misawa, Takahiro and Yamaji, Youhei},
  journal = {Phys. Rev. Res.},
  volume = {4},
  issue = {2},
  pages = {023177},
  numpages = {19},
  year = {2022},
  month = {Jun},
  publisher = {American Physical Society},
  doi = {10.1103/PhysRevResearch.4.023177},
  url = {https://link.aps.org/doi/10.1103/PhysRevResearch.4.023177}
}

@misc{comsol64,
  title        = {{COMSOL Multiphysics\textregistered\ v.\ 6.4}},
  howpublished = {\url{https://www.comsol.com}},
  author       = {{COMSOL AB}},
  note         = {{Stockholm, Sweden}},
}

% ---------- Supplementary Information appended in SAME PDF ----------
\clearpage
\onecolumngrid            % switch to one-column layout for SI (like the example)
\beginsupplement

\EnableTOC  % <-- SI sections will now be written to the .toc

% SI title page + contents page (roughly like the example)
\begin{center}
  {\Large\bfseries Supplemental Material:\par}
  \vspace{0.8em}
  {\Large\bfseries Observation of ring states in a delicate topological insulator\par}
\end{center}
% \title{Supplementary Information: \\Observation of ring states in a delicate topological insulator}

\vspace{1.5em}

% Now allow ToC entries (SI only)
% \setcounter{tocdepth}{2}
\renewcommand{\contentsname}{CONTENTS}
\addtocontents{toc}{\protect\setcounter{tocdepth}{2}} % enable SI ToC entries
\tableofcontents
\clearpage

\section{Wilson loop phase and hybrid Wannier-center position}
\label{sec:wilson}

Hybrid Wannier functions~\cite{Coh2009} of a band with index $n$ and normalized Bloch eigenvectors $\ket{u_{n,\vec k}}$ are defined as the Fourier transform
\begin{equation*}
    \ket{w_{n,y,k_x}} = \frac{b}{2\pi}\int_0^{2\pi/b} dk_y  e^{ik_y y}|u_{n,\vec k}\rangle.
\end{equation*}
We use an embedding or ``gauge'' convention where $|u_{n,\vec k}\rangle=|u_{n,\vec k + \vec G}\rangle$ for all reciprocal lattice vectors $\vec G$. Their localization in $y$-direction
\begin{equation}
\label{eq:wannier}
    \Bar y(k_x)=\bra{w_{n,y,k_x}} y \ket{w_{n,y,k_x}} \;\;\; ({\rm mod}\, b)
\end{equation}
is linked to the phase of the Wilson loop \cite{Zak1982}. The Wilson loop for a single band $n$ is defined as
\begin{align*}
    \mathcal{W}_n(k_x) = e^{i \lambda_n(k_x)},
\end{align*}
where $\lambda_n(k_x)$ is the Wilson loop phase of band $n$ at a fixed $k_x$
\begin{align}
    \label{eq:wilson_phase}
    \lambda_n(k_x)= \int_0^{2\pi/b} dk_y \bra{u_{n, {\vec k}}}\partial_{k_y}\ket{u_{n, {\vec k}}} \; \; \; ({\rm mod} \,  2\pi).
\end{align}
$\lambda_n(k_x)$ is related to the hybrid Wannier-center position along $y$ in Eq.~(\ref{eq:wannier}) via $\lambda_n(k_x)=2 \pi \Bar y(k_x)/b$ $({\rm mod} \,  2\pi)$~\cite{Coh2009}.

\subsection{Quantization on the mirror-invariant lines in the two-band model}

Our system is invariant under the mirror symmetry $M_x: (x,y)\to (-x,y)$, which sends $(k_x, k_y) \mapsto (-k_x, k_y)$. 
Therefore, the lines $k_x = 0$ and $k_x = \pi/a$ are mirror-invariant.
In the two-orbital $\{s, p\}$ model (see Eq.~(\ref{eq:dynamical_matrix_main}) of the main text) the orbitals have opposite mirror parity, and mirror symmetry enforces vanishing inter-orbital coupling on these lines:
\begin{align*}
    \mathcal f_{sp}(0,k_y)=\mathcal f_{sp}(\pi/a,k_y)=0.
\end{align*}
Consequently, along $k_x=0$ and $k_x=\pi/a$ the bands are fully orbital-polarized, i.e., the lowest band is purely $s$-like along $k_x=0$ and purely $p$-like along $k_x=\pi/a$.
For such a fully orbital-polarized eigenstate one can choose a smooth gauge along $k_y$ with 
\begin{align*}
  \bra{u_{n, {\vec k}}}\partial_{k_y}\ket{u_{n, {\vec k}}}=0, %\; \; \; \vec k=(0,k_y) \; {\rm and} \; \vec k=(\pi/a,k_y)
\end{align*} 
where $\vec k=(0,k_y)$ or $\vec k=(\pi/a,k_y)$.
This implies that on the mirror lines the Wilson loop phase [Eq.~(\ref{eq:wilson_phase})] is quantized as
\begin{align*}
    \lambda_n(k_x=0)=2\pi l, \; \; \; \lambda_n(k_x=\pi/a)=2\pi m, \; \; \; l, m \in \mathbb Z.
\end{align*}
Away from the mirror lines, $f_{sp}(\vec k)$ hybridizes the orbitals and opens a bulk gap.
A delicate topological phase is realized when the Wilson loop phase undergoes a winding between the two mirror-invariant lines, i.e.,
\begin{align*}
    |\lambda(k_x=\pi/a) - \lambda(k_x=0)| = 2\pi|m-l|, \;\;\; l \neq m.
\end{align*}
% $|\lambda(k_x=\pi/a) - \lambda(k_x=0)| = 2\pi|m-l|$, with $l \neq m$
Equivalently, the hybrid Wannier center $\Bar{y}(k_x)$ [Eq.~(\ref{eq:wannier})] shifts by $|m-l|$ lattice constants $b$ as $k_x$ is swept from $0$ to $\pi/a$, which means that, even for the case $|m-l|=1$, it cannot be kept inside a single real-space unit cell for all $k_x$, as illustrated in Fig.~\ref{fig:Fig0}b.
This is the defining multicellularity of a delicate topological insulator~\cite{Nelson2021, Nelson2022, Chen2024}.

\subsection{Loss of quantization in the three-band model}

When the additional mirror-odd orbital is included, even on a mirror-invariant line, the occupied state can involve multiple mirror-odd components, and in general $\bra{u_{n, {\vec k}}}\partial_{k_y}\ket{u_{n, {\vec k}}} \neq 0$.
In this case, the Wilson loop phase of band $n$ is no longer quantized, i.e.,
\begin{align*}
    |\lambda(k_x=\pi/a) - \lambda(k_x=0)| \neq  2\pi |m-l|.
\end{align*}
In our $\{s, p, \tilde p \}$ three-band model, hybridization with the $\tilde p$ orbital leads to a loss of quantization of the lowest band as shown in Fig.~\ref{fig:Fig3}c of the main text.
% As a result, the delicate topology of the two lower bands is trivialized.
% for example three bands, pinned to 0, \epsilon, -\epsilon, still sum to 1

\section{Sample design}
\label{sec:sample_design}

We implement a delicate topological insulator as a perturbative metamaterial~\cite{Matlack18}, i.e., a lattice of weakly interacting mechanical plate resonators whose dynamics can be mapped onto a reduced discrete model in a targeted frequency window.
In this approach, local plate modes provide the effective orbital degrees of freedom, while thin connecting arms generate the desired inter-site couplings.
Holes in plates are used to shift on-site energies by tuning the local resonance frequencies and function as local impurities.

We first specify a target tight-binding model with two orbitals $s$ and $p$ per unit cell that realize the desired delicate topological phase via an $s$-$p$ band inversion and SSH-type couplings along the $y$-direction. The corresponding Bloch dynamical matrix $\mathcal{D}({\vec k})$ is chosen such that the intra-orbital hoppings along $x$ have opposite sign for the two orbitals, while inter-orbital couplings $\mathcal{f}_{sp}({\vec k})$ open a full bulk gap and generate a quantized Wilson-loop phase winding over half the Brillouin zone (see Eq.~(\ref{eq:dynamical_matrix_main}) and Fig.~\ref{fig:Fig0}(a), (b) of the main text).

To translate this discrete model into a physical geometry, each unit cell is realized by millimetre-scale silicon plates.
Two plates define one unit cell, where one of the plates is designed to host a mirror-even ($s$) and the other plate is designed to host a mirror-odd ($p$) in the desired frequency range ($140-\SI{170}{\kilo \hertz}$).
Since we target modes that are almost degenerate the sizes of the $s$ and $p$ plate are almost identical. 
The $p$ plate, however, is slightly larger since the number of nodes of the targeted $p$ mode is higher than for the $s$ plate (see Fig.~\ref{fig:Fig0}(c)--(e) of the main text). 
We remove mass at positions where higher-order modes close in frequency have their maxima.
The plates are connected by a set of thin elliptic arms that weakly hybridize the chosen plate modes into Bloch bands. The position, lengths, and widths of the arms are used as design knobs to control the magnitude and sign of the effective hoppings.
The coupling magnitude is tuned by the arm position relative to the nodal lines.
The trenches around the arms in the design minimize coupling to higher-order modes by bypassing areas where those modes exhibit large displacements~\cite{Serra-Garcia18}.
Following the perturbative-metamaterials methodology, we perform finite-element simulations of small plate arrays using the finite-element method package COMSOL Multiphysics~\cite{comsol64} to extract a reduced-order description of the structure:
plate modes are projected onto a local basis, and a Schrieffer-Wolff-type transformation is used to obtain an effective tight-binding model of the selected modes (\suppls~\ref{sec:effective_model_extraction} and \ref{sec:full_tb_model}).
The plate and arm geometries are then iteratively tuned so that the numerically extracted modes and couplings match the target tight-binding parameters within the weak-coupling regime~\cite{Matlack18}.

We realize four different designs with unit cell lattice constants $a = \SI{3.5}{\milli \meter}$ and $b = \SI{10}{\milli \meter}$ in silicon wafers.
Three wafers host arrays of $15 \times 6$ unit cells: one bulk reference sample without impurities, and two impurity samples each hosting two impurities, where the impurities are implemented as a pair of circular holes of radius $r$ on selected $s$ plates.
The first impurity sample hosts an impurity with $r=\SI{0.25}{\milli \meter}$ and $r=\SI{0.45}{\milli \meter}$, while the second impurity sample hosts an impurity with $r=\SI{0.35}{\milli \meter}$ and $r=\SI{0.55}{\milli \meter}$.
For each impurity sample the impurity plates are located in the center column of unit cells and two unit cells separated in $y$, to both minimize impurity-edge and impurity-impurity coupling.
In addition, the strengths of the impurities located on one sample are chosen such that hybridization between the impurities is minimal.
The fourth wafer contains an array of $21 \times 4$ unit cells designed to support edge states~\cite{Khalaf2021} (\suppl~\ref{sec:edge_states}).

\section{Fabrication}
\label{sec:fabrication}

All samples are fabricated on double-side–polished silicon wafers.
The wafers have a nominal diameter of $\SI{100}{\milli \meter}$ with a tolerance of $\pm \SI{0.5}{\milli \meter}$, a thickness of $\SI{275}{\micro \meter} \pm \SI{10}{\micro \meter}$ and a total thickness variation below $\SI{1}{\micro \meter}$.
The crystal orientation is specified as [100] with a tolerance of $\pm \SI{0.5}{\degree}$. 
For all designs, the in-plane $x$-axis of the lattice is aligned to the wafer flat so that $x$ is parallel to the [110] crystal axis, while the plate normal is defined by the [100] direction.
We process all wafers using the same fabrication sequence.
Prior to structuring, we sputter a $\SI{500}{\nano \meter}$ aluminum layer on the backside of each wafer, which serves as an etch stopping layer during through-wafer deep reactive-ion etching (DRIE).
On the front side, a $\SI{6.2}{\micro \meter}$ thick photoresist layer is spin-coated and patterned to define the lattice. 
Hexamethyldisilazane is used as an adhesion promoter.

The structures are etched by DRIE at $\SI{0}{\celsius}$ using 365 Bosch cycles, which results in complete through-wafer penetration of the patterned areas. 
After etching, the structures are released by first stripping the photoresist in hot dimethyl sulfoxide and subsequently removing the backside aluminum in a hot aluminum etchant, thereby freeing the plates while leaving the surrounding silicon intact.
A $\SI{60}{\nano \meter}$ aluminum film is then sputtered onto the front surface to provide a reflective surface and to facilitate alignment of the laser interferometer with respect to the lattice. 
Finally, each wafer is bonded to a rigid aluminum frame, which clamps the outer perimeter of the patterned region and enforces fixed boundary conditions for the out-of-plane modes along the sample edges.
Each aluminum frame has a slit close to which the piezo is attached on the wafer for excitation.
This slit is located in the center of the right edge for the bulk and impurity samples and in the center of the top and bottom edges for the edge sample.

Deviations between the fabricated samples and the ideal model may arise from several sources. 
Across each wafer, the thickness variation remains below $\SI{1}{\micro \meter}$ (less than $0.4\%$ of the wafer thickness), but may still lead to small shifts in the absolute resonance frequencies of single plates across the wafer. 
The DRIE process produces sidewalls that deviate slightly from vertical.
Measurements at multiple locations across the wafers give sidewall angles between $\SI{1}{\degree}$ and $\SI{2.7}{\degree}$, which modify the effective in-plane dimensions of the plates. 
The in-plane orientation of the patterns with respect to the silicon crystal axes is limited by the accuracy of the maskless aligner ($\pm \SI{0.6}{\degree}$) and the specification of the wafer flat location ($\pm \SI{0.5}{\degree}$).

\section{Signal analysis}
\label{sec:signal_analysis}

All measurements are performed with a lock-in amplifier, while the sample is excited by a thickness-mode vibration piezoelectric actuator (SMD07T05R412WL from STEMiNC).
The drive frequency is swept over the range $\SIrange{100}{180}{\kilo\hertz}$, which is far below the specified resonance frequency of the piezo itself ($\SI{4.25}{\mega\hertz}\pm 5\%$).
To extract the amplitude and phase response of individual plates, the lock-in amplifier receives the real-time out-of-plane displacement signal from an interferometer (IDS3010 from attocube) focused on the chosen measurement points of the individual plates (\suppl~\ref{sec:orbital_resolved_measurements_normalization}).
Our experiment is only sensitive to out-of-plane modes, as desired.

We measure plate displacement amplitudes of up to $\SI{500}{\nano \meter}$ and estimate the uncertainty of these amplitudes to be approximately $\sigma_A \approx \SI{155}{\pico\meter}$, obtained from the variation of the extracted response amplitude over repeated measurements. 
The specified systematic uncertainty of the interferometer (about $\SI{5}{\pico\meter}$) is negligible.
The orbital-resolved density of states in Fig.~\ref{fig:Fig3}(b) of the main text and circle sizes in the spatial mode profiles [Fig.~\ref{fig:Fig2}(b)--(e)] represent the normalized squared displacement amplitudes $\bar A^2$ of the corresponding modes.
We describe the detailed orbital-resolved measurement and normalization procedure in the \suppl~\ref{sec:orbital_resolved_measurements_normalization}.
The propagated uncertainty is $\sigma_{\bar A^2}=2\bar A^2 \sigma_A/A$, where the largest relative uncertainties occur for the smallest amplitudes.
This uncertainty is not plotted explicitly for better readability.

Measurements of the band-structure and density of states (i.e., the data shown in Fig.~\ref{fig:Fig1} and Fig.~\ref{fig:Fig3}(b) of the main text) are generally performed with a frequency resolution of $\SI{60}{\hertz}$ in ambient air, where the measured quality factors are on the order of $10^{3}$.
To resolve the band-gap response, i.e., impurity-induced ring states, impurity states, and edge states, with high spectral resolution (see Fig.~\ref{fig:Fig2} of the main text and Fig.~\ref{fig:Fig4} in the \suppl~\ref{sec:edge_states}), we reduce the pressure in our setup to about $\SI{0.7}{m\bar}$, leading to quality factors of up to $10^{5}$.
Measurements in vacuum are performed with a frequency resolution of down to $\SI{0.5}{\hertz}$.

\section{Orbital-resolved measurements and normalization}
\label{sec:orbital_resolved_measurements_normalization}

\begin{figure}
    \centering
    \includegraphics[width=0.45\columnwidth, clip, trim= 0 20 0 20]{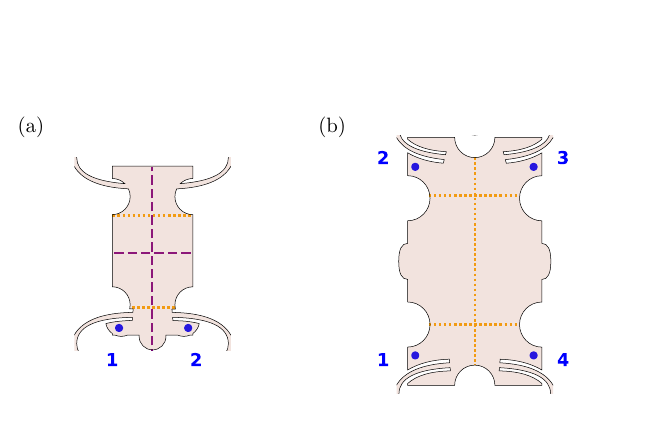}
    \caption{{Measurement points for orbital-resolved readout.}
    (a) $s$-plate geometry. Dotted orange and dashed purple lines indicate nodal lines of the $s$ and $\tilde p$ modes, respectively.
    The response is measured at points $1$ and $2$ (blue dots) and combined to project onto the mirror-even ($s$) and mirror-odd ($\tilde p$) modes (cf. Fig.~\ref{fig:Fig0}d and Fig.~\ref{fig:Fig2}f of the main text).
    (b) $p$-plate geometry. Dotted orange lines indicate nodal lines of the $p$ mode.
    The response is measured at points $1-4$ (blue dots) and combined to isolate the $p$-mode contribution (see Fig.~\ref{fig:Fig0}e).
    }
    \label{fig:test}
\end{figure}

We identify plate modes from the out-of-plane response measured at a small set of points [Fig.~\ref{fig:test}].
At measurement point $j$ we record the amplitude $A_{\gamma, j}(\nu)$ and the phase $\phi_{\gamma, j}(\nu)$ and define the complex response
\begin{equation*}
    a_{\gamma, j}(\nu) = A_{\gamma, j}(\nu) e^{i\phi_{\gamma, j}(\nu)},
\end{equation*}
where $\gamma \in \{ s, p\}$ labels the plate type and $\nu$ denotes the frequency.
For an $s$ plate we use the two points $(1, 2)$ shown in Fig.~\ref{fig:test}(a), chosen at equal distance from the $y$-mirror axis.
The symmetric and antisymmetric combinations distinguish $s$ and $\tilde p$ mode:
\begin{align*}
    A_s(\nu) &= a_{s, 1}(\nu) + a_{s, 2}(\nu) \\ A_{\tilde p}(\nu) &= a_{s, 1}(\nu) - a_{s, 2}(\nu).
\end{align*}
For a $p$ plate we measure at the four points $1-4$ shown in Fig.~\ref{fig:test}(b) and project onto the $p$ mode by computing
\begin{equation*}
    A_p(\nu) = a_{p, 1}(\nu) + a_{p, 2}(\nu) - a_{p, 3}(\nu) - a_{p, 4}(\nu).
\end{equation*}
To quantitatively compare the measured responses of the different modes we need to normalize $A_{s}(\nu)$, $A_{p}(\nu)$, $A_{\tilde{p}}(\nu)$. For the $\alpha=s$ and $\alpha=p$ modes, where we can assume that all their spectral weight lies within our measurement window $[\nu_{\ssm min},\nu_{\ssm max}]=[140\,{\rm kHz},170\,{\rm kHz}]$, we can infer the normalization via the completeness relation 
\begin{equation}
   N_{\alpha}=\int_{\nu_{\ssm min}}^{\nu_{\ssm max}} d\nu |A_{\alpha}(\nu)|^2, \quad \Rightarrow \bar A_{\alpha}(\nu) = \frac{A_{\alpha}(\nu)}{\sqrt{N_{\alpha}}},
\end{equation}
providing us with the normalized wave function amplitudes $\bar A_{\alpha}(\nu)$. For $\alpha=\tilde p$ matters are slighlty more complicated as we do not have reliable data over the full frequency range of the next higher band. We therefore resort to a first principle calculation of the mode shapes in COMSOL Multiphysics~\cite{comsol64} and compare the relative weight of the $s$ and $\tilde p$ modes at the points $(s,1)$ (the weight at $(s,2)$ is related by mirror symmetry and does not yield extra information)
\begin{equation*}
    N_{s \tilde p} = \frac{|w_s(\vec{r}_1)|^2}{\sum_{\vec{r} \in \mathcal{A}} |w_s(\vec{r})|^2} \left(\frac{|w_{\tilde p}(\vec{r}_1)|^2}{\sum_{\vec{r} \in \mathcal{A}} |w_{\tilde p}(\vec{r})|^2}\right)^{-1},
\end{equation*}
where $w_\alpha(\vec{r}_1)$ is the out-of-plane displacement for the mode $\alpha \in \{ s, \tilde p\}$ at the measurement point $\vec{r}_1$ $(s,1)$ on the $s$ plate and $\mathcal{A}$ is the $s$ plate surface area in the $xy$ plane.
We find that 
\begin{equation*}
    N_{s \tilde p}=0.29, % 0.46?
\end{equation*}
and normalize $\bar A_{\tilde p}(\nu)=A_{\tilde p}(\nu)\sqrt{N_{s\tilde p}/N_s}$.

\section{Band-structure measurements}
\label{sec:orbital_resolved_bulk_measurements}

\begin{figure}
    \centering
    \includegraphics{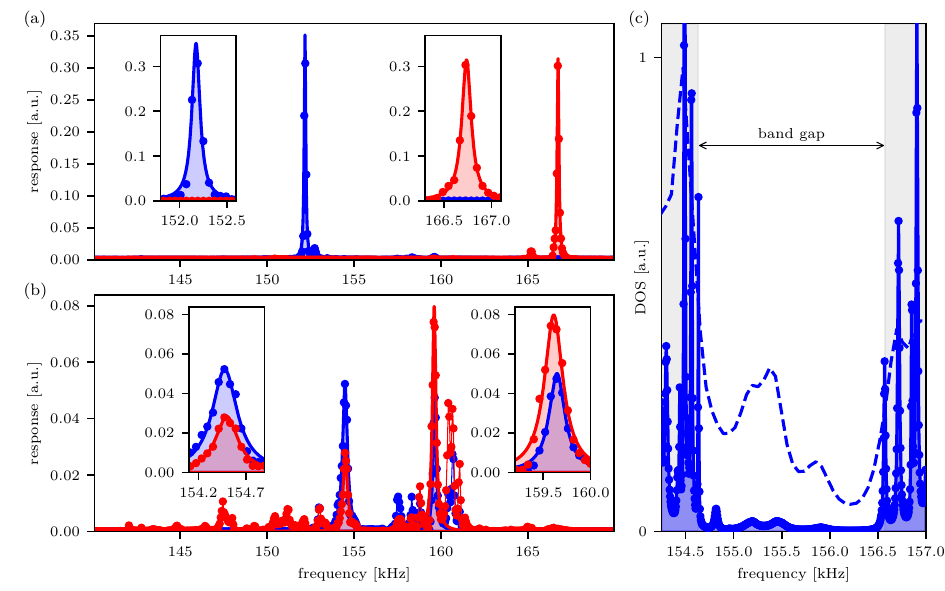}
    \caption{{Orbital-resolved Fourier spectra and extracted polarization along high-symmetry lines.}
    % \textbf{Dispersion analysis from orbital-resolved Fourier spectra.}
    (a), (b)
    Representative orbital-resolved Fourier spectra for two selected $\vec k$ points along the high-symmetry path depicted in Fig.~\ref{fig:Fig0}(a) of the main text, showing the squared response of the $s$ and $p$ mode as blue and red markers, respectively.
    (a) shows the Fourier spectrum at $\Gamma$ ($\vec k =(0,0)$), i.e., on the mirror-invariant line, and (b) shows the Fourier spectrum for a $\vec k$ between $\Gamma$ and $X$ (not on the mirror-invariant line).
    Insets show zooms of the resonances with largest response in each band which are used for computing the polarization (fraction between $s$ and $p$ response at this frequency), see Fig.~\ref{fig:Fig0}(a) of the main text.
    Solid curves in the insets are Lorentzian fits used to extract resonance frequencies and peak amplitudes.
    (c) $s$-mode projected DOS in the frequency window around the bulk gap measured in vacuum ($\sim \SI{0.7}{m\bar}$).
    The dashed curve indicates the same measurement performed in ambient air.
    Grey shaded regions indicate the two bulk bands.
    The range of the band gap is determined from the vacuum DOS data by fitting Lorentzian functions to the DOS peaks associated with the bands, i.e., at frequencies where the DOS begins to rise sharply.
    }
    \label{fig:FigSI1}
\end{figure}

As presented in Fig.~\ref{fig:Fig1}(a) of the main text we characterize the bulk response of the phononic delicate topological insulator by measuring its dispersion along high-symmetry lines in the Brillouin zone.
We first perform orbital-resolved measurements on all plates of the bulk sample by sweeping the piezo frequency from $\SI{140}{\kilo \hertz}$ to $\SI{170}{\kilo \hertz}$ with a frequency resolution of $\SI{60}{\hertz}$ in ambient air.
We then Fourier transform the response and fit the peaks with the highest $s$-mode and $p$-mode intensity in the Fourier transformed signal at each $\vec k$ along the high-symmetry path to Lorentzian functions.
Fig.~\ref{fig:FigSI1}(a) and (b) show the orbital-resolved Fourier transformed signals at two distinct $\vec k$-points in the Brillouin zone: Fig.~\ref{fig:FigSI1}(a) shows the Fourier spectrum at $\Gamma$ and Fig.~\ref{fig:FigSI1}(b) the one at a $\vec k$ point between $\Gamma$ and $X$.
For each band, the polarization at a certain $\vec k$ with respect to $s$ or $p$ mode (shown in Fig.~\ref{fig:Fig1}(a) of the main text) is determined as the fraction between the $s$-mode and $p$-mode response at the frequency of the peak with the largest response.
As expected, the system is orbital-polarized at $\Gamma$ (on the mirror-invariant line), i.e., the lower band is purely $s$-like and the upper band purely $p$-like, see Fig.~\ref{fig:FigSI1}(a).
Away from the mirror-invariant line, the orbitals mix [Fig.~\ref{fig:FigSI1}(b)].

Fig.~\ref{fig:FigSI1}(c) compares the $s$-orbital projected density of states (DOS) measured in vacuum ($\sim \SI{0.7}{m\bar}$) with the same measurement performed in ambient air.
For both measurements we exclude unit cells at the edges and close to the exciter.
From the vacuum DOS we estimate the location of the band gap by fitting Lorentzian functions to the peaks where the DOS increases sharply.
The location of the bulk bands is indicated by the grey shaded areas in Fig.~\ref{fig:FigSI1}(c). 
We observe that, due to the broadening of the resonances in ambient air, the DOS in the estimated region of the band gap is increased, cf. Fig.~\ref{fig:Fig3}(b) of the main text.

\section{Cross-sample frequency alignment}
\label{sec:app_bandstructure_mapping}

\begin{figure*}
    \centering
\includegraphics{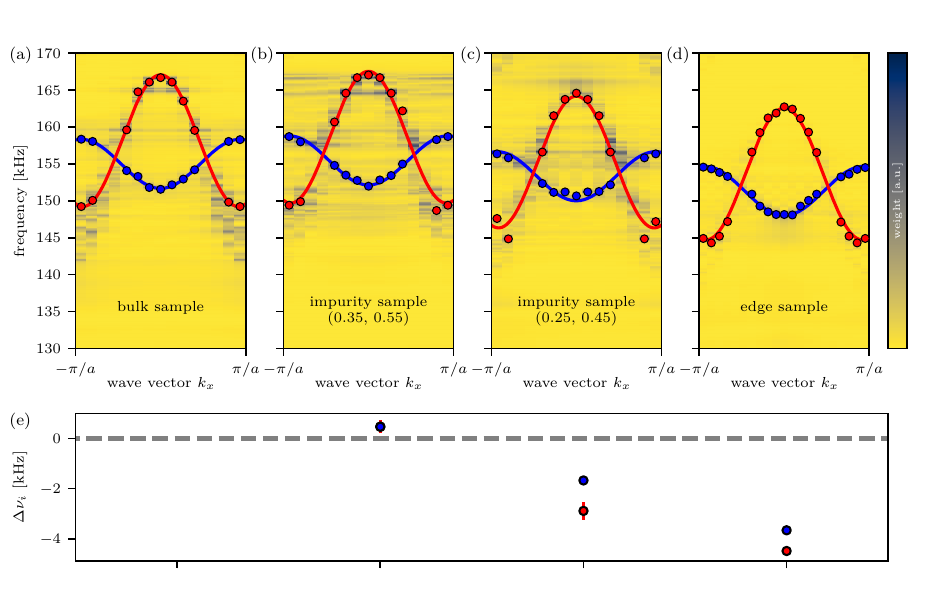}
    \caption{{Cross-sample frequency alignment.} 
    (a)--(d) Spatial Fourier transform of the response along one central row of unit cells for a bulk sample [(a)], a sample with two impurities of radius \SI{0.35}{\milli\meter} and \SI{0.55}{\milli\meter} [(b)], a sample with two impurities of radius \SI{0.25}{\milli\meter} and \SI{0.45}{\milli\meter} [(c)], and an edge sample [(d)]. 
    The color scale shows the Fourier weight (arbitrary units). 
    Red and blue circles indicate, for each wave vector $k_x$, the frequency of the maximum Fourier amplitude of the $s$- and $p$-plate responses, respectively. 
    These points are fitted with cosine dispersions [Eq.~(\ref{eq:dispersion_bulk})]. 
    For the impurity and edge samples, the only fit parameter is the global frequency offset, whereas all other cosine dispersion parameters are fixed to the values obtained from the bulk-sample fits [Eq.~(\ref{eq:dispersion_samples})]. 
    (e) Frequency shifts $\Delta\nu_i$ [Eq.~(\ref{eq:dispersion_shift})] of the fitted bands with respect to the corresponding bulk bands, extracted from the fits in (a)--(d).
    The error bars indicate the standard deviation obtained from the fits.
    The horizontal grey dashed line indicates zero frequency shift, i.e., the bulk-sample reference.
    }
    \label{fig:FigSI4}
\end{figure*}

Variations in the wafer thickness of up to $\SI{20}{\micro\meter}$ lead to global shifts of the mechanical resonance frequencies.
To quantitatively compare the measured frequency spectra and band structures of the four samples (bulk sample, two impurity samples, and edge sample), we align their dispersions to a common frequency reference.
The corresponding data analysis is shown in Fig.~\ref{fig:FigSI4}.

For each sample, we first measure the $s$- and $p$-mode responses along a central row of unit cells.
We then perform a spatial Fourier transform of the $s$- and $p$-mode responses separately and, for each wave vector $k_x$, extract the frequency at which the Fourier amplitude is maximal.
The bulk sample is chosen as the reference.
For this sample, we fit the extracted maxima with
\begin{align}
A_{\text{bulk}, \alpha} \cos\!\left(b_{\text{bulk}, \alpha} k_x\right) + \nu_{\text{bulk}, \alpha},
\label{eq:dispersion_bulk}
\end{align}
where $A_{\text{bulk}, \alpha}$, $b_{\text{bulk}, \alpha}$, and $\nu_{\text{bulk}, \alpha}$ are fit parameters and $\alpha \in \{s,p\}$ [Fig.~\ref{fig:FigSI4}{(a)}].

For the impurity and edge samples, we fit the corresponding maxima to the same functional form but keep the band shape fixed to that of the bulk sample, i.e.
\begin{align}
A_{\text{bulk}, \alpha} \cos\!\left(b_{\text{bulk}, \alpha} k_x\right) + \nu_{i, \alpha},
\label{eq:dispersion_samples}
\end{align}
with only the frequency offset $\nu_{i, \alpha}$ as a free parameter [Fig.~\ref{fig:FigSI4}{(b)--(d)}].
From these offsets we obtain the frequency shifts relative to the bulk sample,
\begin{align}
\Delta \nu_{i, \alpha} = \nu_{i, \alpha} - \nu_{\text{bulk}, \alpha},
\label{eq:dispersion_shift}
\end{align}
which are plotted in Fig.~\ref{fig:FigSI4}{(e)}. 
The horizontal dashed line indicates zero shift (bulk reference).
In Fig.~\ref{fig:Fig2} of the main text and Fig.~\ref{fig:Fig4} we compare resonance frequencies and dispersions of different samples. 
We compensate for the sample-dependent frequency shift and plot all data relative to the bulk reference by adding the $s$-dispersion frequency shift $\Delta \nu_{i, s}$ to the impurity and edge sample data since only data extracted from $s$-plate responses is compared.

\section{Impurity spectroscopy}
\label{sec:impurity_spectroscopy}

\begin{figure*}
    \centering
\includegraphics[width=\textwidth]{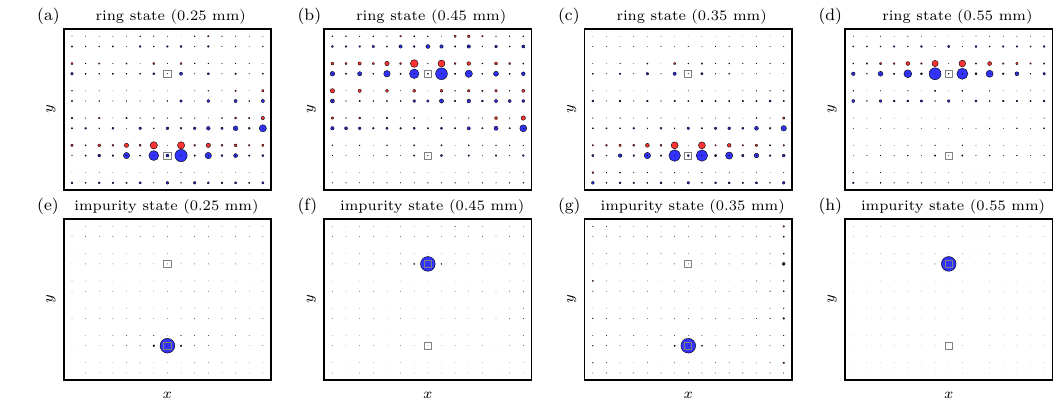}
\caption{{Impurity-induced ring states and impurity states.} {(a)--(d)} Spatial profiles of all four ring states (the corresponding resonance frequencies are plotted in the inset of Fig.~\ref{fig:Fig2}(a) of the main text).
    {(e)--(h)} Spatial profiles of all four $s$ impurity states (the corresponding resonance frequencies are plotted in Fig.~\ref{fig:Fig2}(a) of the main text).
    Each circle area is proportional to the normalized squared amplitude of the $s$ (blue), $\tilde{p}$ (orange), and $p$ (red) modes, and the impurity sites are marked by grey squares.
    One impurity sample hosts impurities with $r=\SI{0.25}{\milli \meter}$ and $r=\SI{0.45}{\milli \meter}$. The corresponding ring states are shown in (a) and (b) and the impurity states are depicted in (e) and (f). 
    The second impurity sample hosts impurities with $r=\SI{0.35}{\milli \meter}$ and $r=\SI{0.55}{\milli \meter}$.
    The observed ring states are shown in (c) and (d) and the corresponding impurity states are depicted in (g) and (h).
    }
    \label{fig:FigSI5}
\end{figure*}

To measure the response of the phononic delicate topological insulator to $s$ impurities (results in Fig.~\ref{fig:Fig2} in the main text) we perform orbital-resolved frequency scans below the first band and in the band gap between the first and second band for each of the two impurity samples.
To achieve a higher spectral resolution, these measurements are performed in vacuum ($\sim \SI{0.7}{m\bar}$) and with a frequency resolution of down to $\SI{0.5}{\hertz}$.
For each sample we record the orbital-resolved response on all plates to reconstruct the spatial profiles of impurity-induced modes and distinguish them from bulk modes.
To identify the frequency of impurity-state resonances we fit Lorentzians to the response on the impurity site.
The $s$ mode ring-state and $\tilde p$ mode in-gap resonance frequencies are determined by fitting Lorentzians to the response of one of the neighboring $s$ plates.
All resonance frequency standard deviations obtained from these fits are negligible compared to the standard deviations of the sample-dependent frequency shifts [Fig.~\ref{fig:FigSI4}] which are shown as error bars in Fig.~\ref{fig:Fig2}(a).
% The orbital-resolved responses and spatial profiles for the impurity sample hosting the $\SI{0.25}{\milli \meter}$ and the $\SI{0.45}{\milli \meter}$ impurity are shown in Fig.~\ref{fig:FigSI5}.
All orbital-resolved spatial profiles of the $s$-impurity induced ring and impurity states are shown in Fig.~\ref{fig:FigSI5}.

\section{Boundary obstruction and edge states}
\label{sec:edge_states}

\begin{figure}
    \centering
    \includegraphics{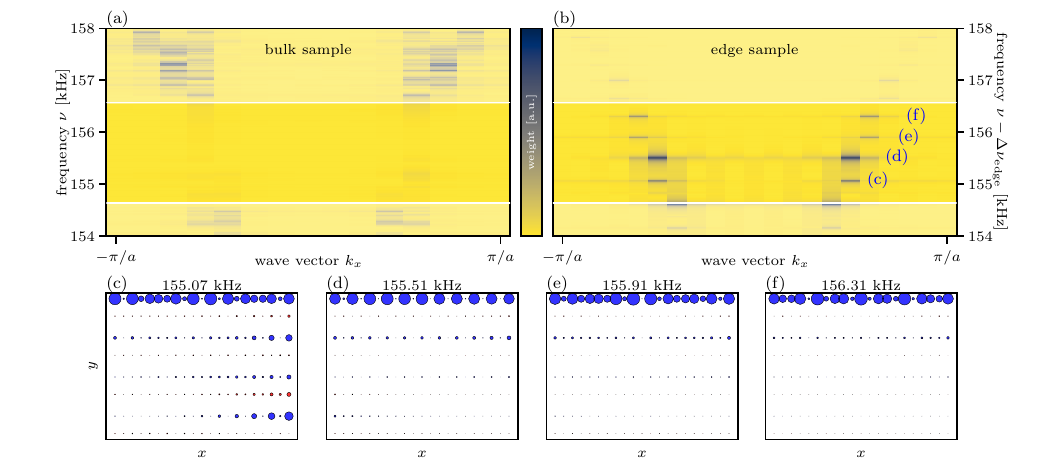}
    \caption{{Edge states.} (a) Spatial Fourier transform of the response of a bulk sample along the upper edge. The white shaded areas indicate the frequency range of the bulk bands (\suppl~\ref{sec:orbital_resolved_bulk_measurements}). The signal is substantially suppressed in the band gap. (b)  The same measurement in a sample where we expect edge states. We observe distinct modes at four frequencies inside the band gap.
    All frequencies are plotted with respect to the bulk sample in $(a)$ by compensating the overall frequency shift $\Delta\nu_{\rm edge}$  of the edge sample (\suppl~\ref{sec:app_bandstructure_mapping}). (c)--(f) Spatial profiles of the modes shown in panel (b). The size of the circles indicates the normalized squared amplitude, the color denotes the mode (blue for $s$ and red for $p$). 
    }
    \label{fig:Fig4}
\end{figure}

Because our system is built from coupled SSH chains along $y$, it inherits a boundary obstruction closely analogous to the SSH case~\cite{Su79,Khalaf2021}: 
for our choice of unit cell (see Fig.~\ref{fig:Fig0}c of the main text; \suppl~\ref{sec:full_tb_model}), edge terminations that intersect the unit cells remain gapped, while edge terminations that cut between unit cells must host in-gap boundary modes.
In the effective $s$/$p$ two-band description, this boundary obstruction is encoded in the Wilson-loop phase winding over half of the Brillouin zone [Fig.~\ref{fig:Fig0}b] and manifests itself as edge modes along the edges orthogonal to the protecting mirror for terminations that cut through Wannier centers.

To study this boundary obstruction experimentally, also in the presence of the $\tilde p$ orbital, we fabricate an additional edge sample with $21 \times 4$ unit cells and, in contrast to the bulk and impurity samples, cut between the unit cells at the bottom and top boundary.
In this configuration we expect edge states along the $x$ edges.
We record the spatial Fourier transform of the response of each $x$-directed edge for the edge sample and compare it to the identical measurement for the bulk sample. 
The results for the upper edge are shown in Fig.~\ref{fig:Fig4}.
For the edge sample, we observe four different modes inside the band gap [Fig.~\ref{fig:Fig4}(b)], which are not present in the bulk sample [Fig.~\ref{fig:Fig4}(a)].
Reconstructed mode profiles confirm that these modes are localized on edge plates, with increased bulk leakage only near the band edges [Fig.~\ref{fig:Fig4}(c)--(f)].

Importantly, we observe these edge states even in the presence of the weakly hybridizing third orbital that trivializes the delicate topology of the two-band subspace.
Their existence shows that the boundary obstruction and associated edge modes are controlled by the underlying even--odd band inversion, rather than by the delicate two-band topology alone.
The observation of ring states (cf. Fig.~\ref{fig:Fig2} of the main text) in such a boundary-obstructed system provides evidence that strong local impurities are powerful probes of bulk properties~\cite{Diop2020}.
Ring states can track band inversion even in situations where the existence of (robust) edge states depends on the choice of unit cell or edge termination.

\section{Effective model extraction}
\label{sec:effective_model_extraction}

We extract a tight-binding model from a finite-element simulation of our structure with unit cells shown in Fig.~\ref{fig:Fig0}(c) using COMSOL Multiphysics~\cite{comsol64}. 
In the framework of perturbative metamaterials \cite{Matlack18}, we solve the two-plate model with periodic boundary conditions at $\Gamma$ ($k_x=k_y=0$). 
From this we extract the eigenmodes and use them as proxies to the Wannier functions. 
We define the eigenmodes with mode number $m$ as $\varphi_\beta^m(\boldsymbol r_i^\gamma)$, where $\beta=u,v,w$ denote the three displacement fields per point $\boldsymbol r_i^\gamma$.
We choose nine representative points $\boldsymbol r_i^\gamma$ per plate $\gamma=s,p$ (labeled by the index $i=1,\dots,9$) at which we evaluate the solutions.
Moreover, we assume that the modes $m$ on the two plates $\gamma$ are well separated in frequency. 
Hence, for each mode we can assign one of the two plates $\gamma_m$: 
\begin{align*}
w_\beta^m(\boldsymbol r_i^\gamma) = \begin{cases} \varphi_\beta^m(\boldsymbol r_i^\gamma) & \text{if}\;\gamma=\gamma_{m} \\ 0 &{\rm otherwise.}\end{cases}
\end{align*}
These resulting plate-centered modes $w_\beta^m(\boldsymbol r_i^\gamma)$ are the Wannier function proxies we use for our analysis. 
Note that this approach is a slight change compared to the original formulation of the perturbative metamaterial technique~\cite{Matlack18, Serra-Garcia18}. 
We believe it yields better results under two conditions: 
First, when the couplers change the single-plate modes significantly, the presented starting point seems better than using single plate modes. 
Second, this only works if you have a structural symmetry, like $M_x$ here, that helps to have these $\Gamma$-point modes reflect well an ``atomic Wannier function". 

Next, we simulate a periodic supercell with $N=L_x \times L_y$ unit cells and project the so-obtained modes onto the calculated Wannier functions.
To obtain an effective two- and three-band tight-binding model we solve a system with $N=5\times 3$ unit cells with periodic boundary conditions.
Note that with this supercell dimensions we can reliably resolve next-nearest neighbor couplings in the $x$-direction and intra-unit cell and nearest neighbor couplings in the $y$ direction. 
Each supercell eigenmode $n$ provides displacement fields
\begin{align*}
    \psi_\beta^{n}(\boldsymbol r_i^\gamma; \vec R),
\end{align*}
sampled at the same relative point pattern $\boldsymbol r_i^\gamma$ in every unit cell $\vec R = r \vec a_x + s \vec a_y$, where $r \in \{1, 2, 3, 4, 5\}$, $s \in \{1, 2, 3\}$ and ${\vec a}_x$ and ${\vec a}_y$ represent the lattice vectors in $x$ and $y$ direction, respectively.
We then compute the overlaps between each supercell eigenmode and each localized orbital translated to each unit cell
% We can now calculate the overlaps 
\begin{align*}
\zeta_{n,[m,\vec R]} = \frac{\sum_{\boldsymbol r_i^\gamma} \sum_{\beta} \psi_\beta^{n}(\boldsymbol r_i^\gamma; \vec R)  w^m_{\beta}(\boldsymbol r_i^\gamma)}{\sum_{\boldsymbol r_i^\gamma} \sum_{\beta} \left[w^m_{\beta}(\boldsymbol r_i^\gamma)\right]^2}.
\end{align*}
$\zeta_{n,[m,\vec R]}$ quantifies how much the single-plate mode $m$ at site $\vec R$ participates in the mode $n$ of the coupled system. 
The square bracket around $[m,\vec R]$ indicates that we think of this combination as one index over which we sum and that it has the same cardinality as $n$. 
Let $\Omega$ be the diagonal matrix of supercell eigenfrequencies $\omega_n$ for the selected modes.
We obtain a real-space tight-binding dynamical matrix $K$ by computing
\begin{align*}
    K = \zeta^\dagger \Omega^2 \zeta.
\end{align*}
From this matrix we subsequently extract short-range hoppings $t_{ij}^{\alpha \beta}$ to define the dynamical matrix $\mathcal D(\vec k)$ in momentum space.

\section{Full tight-binding model and parameters}
\label{sec:full_tb_model}

We write the three-orbital tight-binding dynamical matrix (in units of $\omega^2$) in the basis $\{s, p, \tilde p\}$ as
\begin{equation}
    \mathcal D = \begin{bmatrix}
    \mathcal s({\vec k}) & \mathcal f_{sp}({\vec k})& \mathcal f_{s\tilde p}(\vec k)\\
    \mathcal f_{sp}^{*}({\vec k}) & \mathcal p({\vec k})& \mathcal f_{p\tilde p}(\vec k)\\
    \mathcal f^*_{s \tilde p}({\vec k}) & \mathcal f_{p \tilde p}^{*}({\vec k})& \mathcal {\tilde p}(\vec k) 
    \end{bmatrix}
    \label{eq:dynamical_matrix}
\end{equation}
and parameterize each matrix element by frequencies $\omega_{\alpha}$ and hopping amplitudes $t_{\alpha \beta}^{ij}$, where $\alpha, \beta \in \{s, p, \tilde p \}$. The index $(i, j)$ denotes a displacement by $i {\vec a}_x + j{\vec a}_y$ between unit cells, where ${\vec a}_x$ and ${\vec a}_y$ denote the lattice vectors in $x$ and $y$ direction, respectively.
Mirror symmetry $M_x$ enforces that diagonal terms are even in $k_x$, while couplings between mirror-even and mirror-odd orbitals are odd in $k_x$ and therefore appear with $\sin(k_x)$.
The diagonal dispersions are
\begin{align}
    \label{eq:dispersion_1}
    \mathcal s({\vec k}) & = 
    \omega_{s}^2+ 2t_{ss}^{10} \cos(k_x) +  2t_{ss}^{20}\cos(2k_x),\\
    \mathcal p({\vec k})  &= 
    \omega_{p}^2+ 2t_{pp}^{10} \cos(k_x)+  2t_{pp}^{20}\cos(2k_x) + 2 t_{pp}^{11} \cos(k_x+k_y)
    +2 t_{pp}^{21} \cos(2k_x+k_y) + 2 t_{pp}^{01} \cos(k_y),\\
    \mathcal {\tilde p}(\vec k) &= 
    \omega_{\tilde p}^2+2t_{\tilde p \tilde p}^{10} \cos(k_x)+ 2t_{\tilde p \tilde p}^{20} \cos(2k_x)+2 t_{\tilde p \tilde p}^{11} \cos(k_x+k_y)+2t_{\tilde p \tilde p}^{21} \cos(2k_x+k_y)+2t_{\tilde p \tilde p}^{01} \cos(k_y).
\end{align}
The off-diagonal coupling elements are given as
\begin{align}
    \mathcal f_{sp}({\vec k})&=2 {\rm i} t_{sp}^{10} \sin(k_x) + 2 {\rm i} t_{sp}^{11\uparrow} \sin(k_x)e^{{\rm i}k_y},\\
    \mathcal f_{s\tilde p}({\vec k}) &= 2{\rm i}t_{s\tilde p}^{10}\sin(k_x)+2{\rm i}t_{s\tilde p}^{20}\sin(2k_x)+4{\rm i}t_{s\tilde p}^{11}\sin(k_x)\cos(k_y),\\
    \mathcal f_{p\tilde p}({\vec k})&= t_{p\tilde p}^{00}+ t_{p\tilde p}^{01\downarrow}e^{-{\rm i}k_y} + 2t_{p\tilde p}^{10}\cos(k_x) + 2t_{p\tilde p}^{11\downarrow}\cos(k_x)e^{-{\rm i}k_y}+2t_{p\tilde p}^{20}\cos(2k_x).
    \label{eq:dispersion_6}
\end{align}
Here, the arrows $\uparrow$ and $\downarrow$ indicate unidirectional coupling to unit cells above and below in $y$-direction, respectively, and we set the lattice constants to $a=b=1$ in Eqs.~(\ref{eq:dispersion_1})--(\ref{eq:dispersion_6}).
We consider terms up to $|i| \leq 2$, $|j| \leq 1$ and extract the tight-binding parameters using the procedure described in \suppl~\ref{sec:effective_model_extraction}.
All parameters for both the two-band [Eq.~(\ref{eq:dynamical_matrix_main}) of the main text] and three-band tight-binding model are listed in Tab.~\ref{app:hoppings}.
The corresponding orbital-resolved band structure for the three-band model is shown in Fig.~\ref{fig:Fig3}(a) of the main text.

\begin{table}
  
  % \begin{tabular*}{0.4\columnwidth}{@{\extracolsep{\fill}}
  %                                    R{0.1\columnwidth}
  %                                    L{0.1\columnwidth}
  %                                    R{0.1\columnwidth}
  %                                    R{0.1\columnwidth}}
  \centering
  \begin{tabular*}{0.35\columnwidth}{@{\extracolsep{\fill}} llrr}
\toprule
     &  & two bands& three bands \\
\hline
     $\omega_s$ &$[{\rm kHz}]$ & $152.5$ & $152.7$ \\
    $\omega_p$ &$[{\rm kHz}]$ & $153.6$ & $156.2$ \\
    $\omega_{\tilde p}$ &$[{\rm kHz}]$ &  & $197.4$ \\
    $t_{ss}^{01}$ &$[{\rm kHz}^2]$ & $17.7$ & $18.1$ \\
    $t_{ss}^{10}$ &$[{\rm kHz}^2]$ & $-485.2$ & $-503.8$ \\
    $t_{ss}^{11}$ &$[{\rm kHz}^2]$ & $14.3$ & $11.0$ \\
    $t_{ss}^{20}$ &$[{\rm kHz}^2]$ & $-48.5$ & $-64.6$ \\
    $t_{pp}^{01}$ &$[{\rm kHz}^2]$ & $272.0$ & $-120.3$ \\
    $t_{pp}^{10}$ &$[{\rm kHz}^2]$ & $1678.8$ & $1304.9$ \\
    $t_{pp}^{11}$ &$[{\rm kHz}^2]$ & $-147.3$ & $46.1$ \\
    $t_{pp}^{20}$ &$[{\rm kHz}^2]$ & $-29.5$ & $-27.1$ \\
    $t_{pp}^{21}$ &$[{\rm kHz}^2]$ & $25.3$ & $29.7$ \\
    $t_{sp}^{10}$ &$[{\rm kHz}^2]$ & \color{red}$27.1$ & \color{red}$142.6$ \\
    $t_{sp}^{11}$ &$[{\rm kHz}^2]$ & \color{red}$233.8$ & \color{red}$115.5$ \\
    $t_{\tilde p\tilde p}^{01}$ &$[{\rm kHz}^2]$ &  & $219.5$ \\
    $t_{\tilde p\tilde p}^{10}$ &$[{\rm kHz}^2]$ &  & $7461.6$ \\
    $t_{\tilde p\tilde p}^{11}$ &$[{\rm kHz}^2]$ &  & $32.9$ \\
    $t_{\tilde p\tilde p}^{20}$ &$[{\rm kHz}^2]$ &  & $1680.6$ \\
    $t_{\tilde p\tilde p}^{21}$ &$[{\rm kHz}^2]$ &  & $-135.5$ \\
    $t_{p\tilde p}^{00}$ &$[{\rm kHz}^2]$ &  & $1782.4$ \\
    $t_{p\tilde p}^{01\downarrow}$ &$[{\rm kHz}^2]$ &  & $-2011.5$ \\
    $t_{p\tilde p}^{10}$ &$[{\rm kHz}^2]$ &  & $-468.0$ \\
    $t_{p\tilde p}^{11\downarrow}$ &$[{\rm kHz}^2]$ &  & $261.3$ \\
    $t_{p\tilde p}^{20}$ &$[{\rm kHz}^2]$ &  & $-471.1$ \\
    $t_{p\tilde p}^{21\downarrow}$ &$[{\rm kHz}^2]$ &  & $84.9$ \\
    $t_{s\tilde p}^{10}$ &$[{\rm kHz}^2]$ &  & \color{red}$619.0$ \\
    $t_{s\tilde p}^{11}$ &$[{\rm kHz}^2]$ &  & \color{red}$-62.0$ \\
    $t_{s\tilde p}^{20}$ &$[{\rm kHz}^2]$ &  & \color{red}$36.1$ \\
  \end{tabular*}
  \caption{{Extracted tight-binding parameters}. The tight-binding parameters extracted from a finite-element simulation of a $L_x \times L_y = 5 \times 3$ system for a two-band and three-band model. 
  The corresponding dynamical matrices $\mathcal{D}({\vec k})$ are given in Eq.~(\ref{eq:dynamical_matrix_main}) of the main text and Eq.~(\ref{eq:dynamical_matrix}). The third column gives the values when the numerical solution is projected into a subspace with two orbitals ($s$ and $p$). The fourth column shows the parameters for a three-band model with an additional auxiliary $\tilde p$ orbital on the $s$ plate. 
  Values shown in red denote hopping amplitudes for couplings between mirror-even and mirror-odd orbitals.
  % Due to mirror symmetry, these hopping amplitudes have opposite signs for positive and negative $x$-direction.
  Due to mirror symmetry, the corresponding coupling terms in the Hamiltonian are $\propto 2{\rm i}\sin(k_x)$ rather than $\propto 2\cos(k_x)$.
  }
  \label{app:hoppings}
\end{table}

\section{Ring states and impurity-projected Green's function numerics}
\label{sec:greens_function_numerics}

We now discuss how adding a local impurity to our tight-binding model induces ring states.
First, we discuss the case where a local impurity acts only on one orbital (rank-1) in the $s$/$p$ two-band model.
We then move to a local impurity which acts on two orbitals (rank-2) in the three-band model.
As we described in the main text, the latter case is particularly relevant for our system, since the impurity on the $s$ plate affects both $s$ and $\tilde p$ orbital.
The generalization to higher-rank impurities is straightforward~\cite{Queiroz2024}.

Our clean periodic system is described by the dynamical matrix $\mathcal{D}({\vec k})$ which is given by the $2 \times 2$ matrix in Eq.~(\ref{eq:dynamical_matrix_main}) for the two-band model and by the $3 \times 3$ matrix in Eq.~(\ref{eq:dynamical_matrix}) for the three-band model.
The respective parameters are listed in Tab.~\ref{app:hoppings}.
For each ${\vec k}$ we can solve 
\begin{align*}
    \mathcal{D}({\vec k}) \ket{u_{n, {\vec k}}} = \epsilon_n({\vec k}) \ket{u_{n, {\vec k}}},
\end{align*}
where $\ket{u_{n, {\vec k}}}$ are eigenvectors, $\epsilon_n({\vec k}) = \omega_n^2({\vec k})$ are the eigenvalues (squared frequencies $\omega_n({\vec k})$) and
$n$ is the band index, e.g., $n \in \{1, 2, 3 \}$ for the three-band model.
We define the retarded Green's function for the  system without impurities in the Bloch basis as 
\begin{align*}
    G_0(\epsilon) = \sum_{n, {\vec k}} \frac{\ket{u_{n, {\vec k}}} \bra{u_{n, {\vec k}}}}{\epsilon - \epsilon_n({\vec k})+i 0^+}.
\end{align*}

\subsection{Rank-1 impurity in the two-band model}

\begin{figure}
    \centering
    \includegraphics{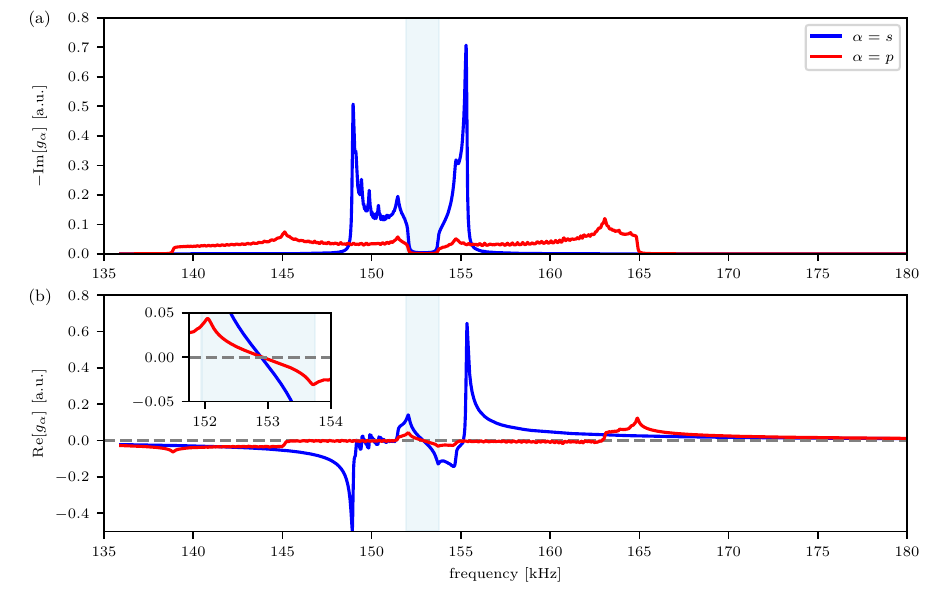}
    \caption{{Rank-1 impurity-projected Green's function in the two-band tight-binding model.} 
    (a) Negative imaginary part of the numerically computed impurity-projected Green's function $g_\alpha$ (see Eq.~(\ref{eq:imp_projected_G_numerics})) versus frequency for the two orbitals $\alpha=s$ (blue) and $\alpha=p$ (red).
    $-{\rm Im} [g_\alpha]$ is directly proportional to the orbital-projected DOS.
    (b) Real part of the numerically computed impurity-projected Green's function $g_\alpha$ versus frequency.
    The inset shows a zoom into the band gap which is indicated by the light blue shaded area.
    The dashed line indicates ${\rm Re}[g_\alpha]=0$.
    }
    \label{fig:SI_greens_function_TB_2band}
\end{figure}

In our $s$/$p$ two-band model [Eq.~(\ref{eq:dynamical_matrix_main}) of the main text] we add a local impurity of strength $U_\alpha$ acting on a single orbital $\alpha \in \{ s, p\}$ (rank-1) at the origin.
The impurity is described by the potential
\begin{align}
    V = U_\alpha \ket{\alpha, \mathbf{0}} \bra{\alpha, \mathbf{0}} =: U_\alpha \ket{\alpha} \bra{\alpha}.
    \label{eq:impurity_potential}
\end{align}
For completeness, we also discuss $\alpha=p$ although we did not implement $p$ impurities in our experiment.
The impurity-projected Green's function is defined as
\begin{align}
    g_\alpha(\epsilon) = \bra{\alpha} G_0(\epsilon) \ket{\alpha} = \sum_{n, {\vec k}} \frac{|\langle \alpha | u_{n, {\vec k}} \rangle|^2}{\epsilon - \epsilon_n({\vec k})+i 0^+} = \frac{1}{N_{\vec k}}\sum_{n, \mathbf{k}} \frac{|u_{n, \alpha}({\vec k})|^2}{\epsilon - \epsilon_n({\vec k})+i 0^+},
    \label{eq:imp_projected_G}
\end{align}
where $u_{n, \alpha}({\vec k}) = \sqrt{N_{\vec k}} \langle \alpha | u_{n, {\vec k}} \rangle$ is the $\alpha$-component of the eigenvector $\ket{u_{n, {\vec k}}}$ with $\alpha \in \{ s, p\}$ and $N_{\vec k}$ is a normalization factor.
The imaginary part of Eq.~(\ref{eq:imp_projected_G}) is directly proportional to the orbital-projected DOS:
\begin{align*}
    {\rm Im} [g_\alpha(\epsilon)] = -\pi \sum_{n, {\vec k}} |u_{n, \alpha}({\vec k})|^2 \delta(\epsilon - \epsilon_n({\vec k})).
\end{align*}
The full Green's function of the perturbed system can now be written in terms of the $T$-matrix
\begin{align*}
    G(\epsilon) = G_0(\epsilon)  + G_0(\epsilon)  T(\epsilon)  G_0(\epsilon) 
\end{align*}
with
\begin{align}
        T(\epsilon) = \frac{V}{1-G_0(\epsilon) V}.
        \label{eq:T_matrix}
\end{align}
Poles of $G(\epsilon)$ and equivalently $T(\epsilon)$  correspond to bound states.
Given Eqs.~(\ref{eq:impurity_potential}) and (\ref{eq:T_matrix}), the only non-zero matrix element of $T(\epsilon)$ is
\begin{align*}
    t_\alpha(\epsilon) = \bra{\alpha} T(\epsilon) \ket{\alpha} = \frac{U_\alpha}{1-U_\alpha g_\alpha(\epsilon)}
\end{align*}
such that the condition for a pole at energy $\epsilon_{\rm b}$ is
\begin{align}
    g_\alpha(\epsilon_{\rm b}) = \frac{1}{U_\alpha}.
    \label{eq:bound_state_condition}
\end{align}
In the strong-impurity limit $|U_\alpha| \rightarrow \infty$, Eq.~(\ref{eq:bound_state_condition}) forces the bound-state energy to approach a zero of the impurity-projected Green's function
\begin{align*}
    g_\alpha(\epsilon_{\rm b}^{*}) = 0.
\end{align*}
Since ${\rm Im} [g_\alpha] = 0$ in the bulk gap, $\epsilon_{\rm b}^{*}$ corresponds to a zero of ${\rm Re} [g_\alpha]$.
In-gap zeros of ${\rm Re} [g_\alpha]$ can be interpreted as spectral attractors: sufficiently strong impurities lead to a pinning of bound states at $\epsilon_{\rm b}^{*}$~\cite{Queiroz2024}.
These pinned in-gap bound states are ring states which are orthogonal to the impurity state $\ket{\alpha}$: they have negligible amplitude on the impurity site but a maximum on a ring of nearby sites.

Since ${\rm Re} [g_\alpha] \rightarrow 0$ also for $|\epsilon| \rightarrow \infty$ there always exists exactly one bound state outside the full spectrum (below the lowest or above the highest band) for each $U_\alpha$~\cite{Queiroz2024}.
This state corresponds to an eigenmode localized predominantly on the impurity site whose frequency is pushed down or up depending on the sign of the local impurity potential. 

We numerically compute the impurity-projected Green's function for our two-band model to study the existence of ring states.
We evaluate Eq.~(\ref{eq:imp_projected_G}) on a discrete grid of $\epsilon$ and replace $0^+$ by a small but finite broadening $\eta$ which we choose as $5\%$ of the frequency range of the lower band gap.
We define a Brillouin zone mesh ${\vec k} = (k_x, k_y)$ with $k_x, k_y \in [-\pi, \pi)$ and with total grid size $N_{\vec k}=N_{k_x} \times N_{k_y} = 200 \times 200$.
For each ${\vec k}$ we diagonalize $\mathcal{D}( {\vec k})$ and compute the weights $|u_{n, \alpha}({\vec k})|^2$.
Finally, for each $\epsilon_i$ we compute
\begin{align}
    g_\alpha(\epsilon_i) = \frac{1}{N_{\vec k}} \sum_{n, {\vec k}} \frac{|u_{n, \alpha}({\vec k})|^2}{\epsilon_i - \epsilon_n({\vec k})+i \eta}.
    \label{eq:imp_projected_G_numerics}
\end{align}
Fig.~\ref{fig:SI_greens_function_TB_2band}(a) shows $-{\rm Im} [g_\alpha]$ (proportional to the $\alpha$-projected DOS) for $\alpha = s, p$.
The $x$ axis is the frequency computed as $\sqrt{\epsilon_i}$.
The shaded region indicates the full band gap between the two bands. 
As expected, both bands have overlap with the $s$ and $p$ orbital, where the $p$-projected DOS is extended over a larger frequency range.
Fig.~\ref{fig:SI_greens_function_TB_2band}(b) displays ${\rm Re} [g_\alpha]$ for $\alpha = s, p$.
Both ${\rm Re} [g_s]$ and ${\rm Re} [g_p]$ cross the zero line in the gap (inset).
This implies that both $s$ and $p$ impurities induce ring states whose energy is pinned to the zero-crossing energies $\epsilon_{{\rm b}, s}^{*}$ and $\epsilon_{{\rm b}, p}^{*}$ inside the band gap for sufficiently strong impurities.

\subsection{Rank-2 impurity in the three-band model}

\begin{figure}
    \centering
    \includegraphics{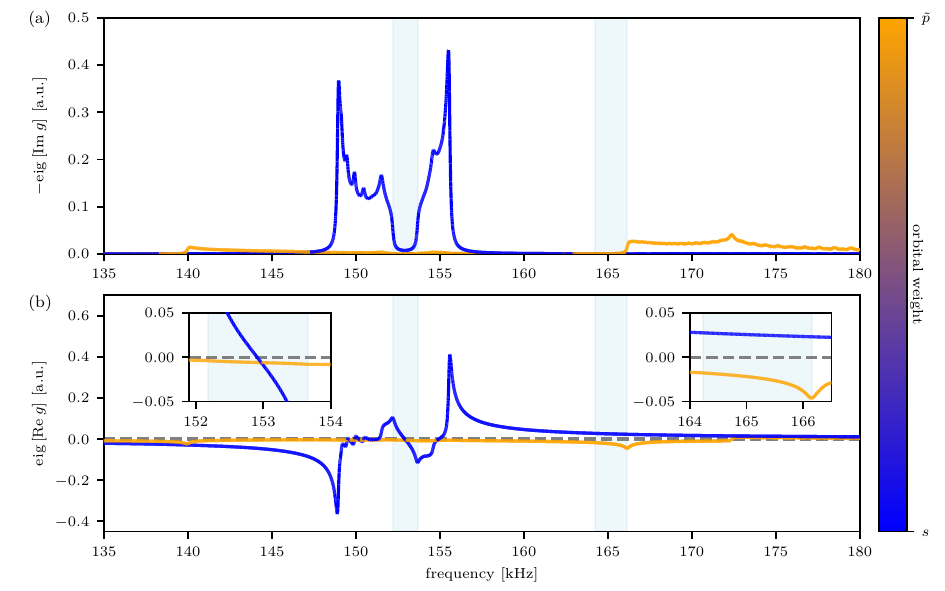}
    \caption{{Rank-2 impurity-projected Green's function in the three-band tight-binding model.} 
    (a) Eigenvalues of the negative imaginary part of the numerically computed impurity-projected Green's function $g$ (see Eq.~(\ref{eq:imp_projected_G_numerics_rank2_numerics})) versus frequency.
    The colors indicate the orbital weight computed as the overlap of the corresponding eigenvector $\ket{v_{{\rm imag}}}$ of $-{\rm Im} \, g$ with the $s$ orbital $|\langle s | v_{{\rm imag}} \rangle|^2$.
    Blue corresponds to $|\langle s | v_{{\rm imag}} \rangle|^2=1$, whereas orange corresponds to $|\langle s | v_{{\rm imag}} \rangle|^2=0$ and therefore maximal overlap with the $\tilde p$ orbital.
    (b) Eigenvalues of the real part of the numerically computed impurity-projected Green's function $g$ versus frequency.
    Again, the colors indicate the orbital weight computed as the overlap of the corresponding eigenvector $\ket{v_{{\rm real}}}$ of ${\rm Re} \, g$ with the $s$ orbital $|\langle s | v_{{\rm real}} \rangle|^2$.
    Blue corresponds to $|\langle s | v_{{\rm real}} \rangle|^2=1$, whereas orange corresponds to $|\langle s | v_{{\rm real}} \rangle|^2=0$.
    The insets show zooms into the two band gaps which are indicated by the light blue shaded areas.
    The dashed line indicates ${\rm eig}[{\rm Re}\,g]=0$.
    }
    \label{fig:SI_greens_function_TB_3band}
\end{figure}

In our physical system, the impurity on an $s$ plate affects both the $s$ and the $\tilde p$ local resonances and is therefore rank-2.
In the $\{ s, \tilde p\}$ basis it reads
\begin{align*}
V =
\begin{bmatrix}
    U_{ss} & 0  \\
    0 & U_{\tilde p \tilde p}
    \end{bmatrix}.
\end{align*}
We define the projected Green's-function matrix in the $\{ s, \tilde p\}$ subspace
\begin{align}
    g(\epsilon) = \begin{bmatrix}
    g_{ss}(\epsilon) & g_{s\tilde p}(\epsilon)  \\
    g_{\tilde ps}(\epsilon) & g_{\tilde p \tilde p}(\epsilon)
    \end{bmatrix}, \;\;\;\;\; g_{\alpha \beta}(\epsilon) = \frac{1}{N_{\vec k}} \sum_{n, {\vec k}} \frac{u_{n, \alpha}({\vec k}) u^{*}_{n, \beta}({\vec k})}{\epsilon - \epsilon_n({\vec k})+i 0^+}.
    \label{eq:imp_projected_G_numerics_rank2}
\end{align}
where, analogously to the rank-1 impurity case, $u_{n, \alpha/\beta}({\vec k}) = \sqrt{N_{\vec k}} \langle \alpha/\beta | u_{n, {\vec k}} \rangle$ is the $\alpha/\beta$-component of the eigenvector $\ket{u_{n, {\vec k}}}$ with $\alpha, \beta \in \{ s, \tilde p\}$ and $N_{\vec k}$ is a normalization factor.
Bound states correspond to poles of the $T$-matrix which in the higher-rank case satisfy
\begin{align*}
    {\rm det}[\mathrm{I}-V g(\epsilon)]=0.
\end{align*}
In the strong-impurity limit $||V|| \rightarrow \infty$, the condition for a bound state at energy $\epsilon_{\rm b}^{*}$ reads
\begin{align*}
    {\rm det}[g(\epsilon_{\rm b}^{*})]=0.
\end{align*}
Hence, it is satisfied if an eigenvalue of $g(\epsilon)$ is zero for a certain $\epsilon_{\rm b}^{*}$~\cite{Queiroz2024}.
To study the existence of ring states, we numerically evaluate $g(\epsilon)$ and look for in-gap zeros of the eigenvalues of the real part of $g(\epsilon)$ (${\rm Im}\,g(\epsilon)$ vanishes element-wise for $\epsilon$ in the bulk gap).
We evaluate Eq.~(\ref{eq:imp_projected_G_numerics_rank2}) on a discrete grid of $\epsilon$ and replace $0^+$ by a broadening $\eta$ which we again choose to be $5\%$ of the lower band gap frequency range.
Again, we define a Brillouin zone mesh ${\vec k} = (k_x, k_y)$ with $k_x, k_y \in [-\pi, \pi)$ and with total grid size $N_{\vec k}=N_{k_x} \times N_{k_y} = 200 \times 200$.
For each ${\vec k}$ we diagonalize $\mathcal{D}({\vec k})$ and compute the weights $u_{n, \alpha}({\vec k}) u^{*}_{n, \beta}({\vec k})$.
Next, for each $\epsilon_i$ we compute the matrix elements of $g$
\begin{align}
    g_{\alpha \beta}(\epsilon_i) = \frac{1}{N_{\vec k}} \sum_{n, {\vec k}} \frac{u_{n, \alpha}({\vec k}) u^{*}_{n, \beta}({\vec k})}{\epsilon_i - \epsilon_n({\vec k})+i \eta},
    \label{eq:imp_projected_G_numerics_rank2_numerics}
\end{align}
where $\alpha, \beta \in \{ s, \tilde p\}$.
We then evaluate the eigenvalues and eigenvectors of the imaginary and real parts of $g$ for each $\epsilon_i$.

Fig.~\ref{fig:SI_greens_function_TB_3band}(a) shows the negative eigenvalues of the imaginary part of $g$ which reproduce the expected orbital-projected DOS.
In the two band gaps (indicated by the grey shaded areas) the orbital-projected DOS vanishes.
The $s$-projected DOS is peaked around the lower band gap, whereas the $\tilde p$-projected DOS is small in the regions of the two lower bands and maximal above the second band gap.
In Fig.~\ref{fig:SI_greens_function_TB_3band}(b) we plot the eigenvalues of the real part of $g$.
Since the eigenvalues of the imaginary part vanish inside the band gaps, in-gap zero crossings of eigenvalues of the real part imply the existence of ring states.
We observe that the $s$-dominated eigenvalues cross zero inside the lower band gap, whereas the $\tilde p$-dominated eigenvalues stay negative.
Hence, the existence of $s$-impurity induced ring states is guaranteed, whereas in-gap $\tilde p$-impurity induced resonances do not correspond to ring states and are not pinned. 
This numerical result is consistent with our experimental observations presented in the main text (cf. Fig.~\ref{fig:Fig2}).

\subsection{Rank-2 impurity induced states on a finite lattice}

\begin{figure}
    \centering
    \includegraphics{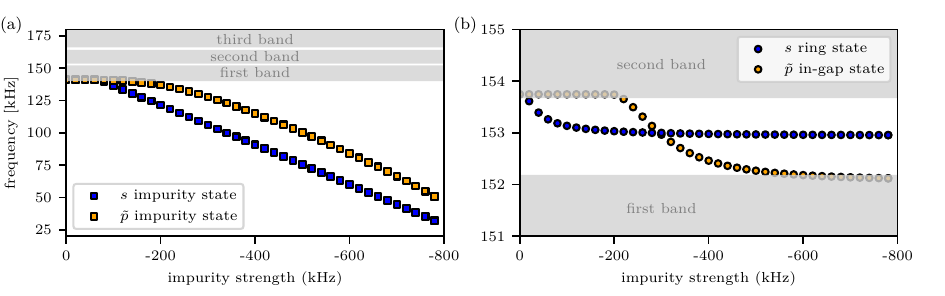}
    \caption{{Impurity-induced ring states in the three-band tight-binding model.}
    Eigenfrequencies of the real-space three-band tight-binding Hamiltonian (parameters in Tab.~\ref{app:hoppings}) for a $L_x\times L_y = 21 \times 21$ lattice with periodic boundary conditions. 
    A single on-site rank-2 impurity affecting $s$ and $\tilde p$ orbital is placed at the lattice center. 
    The impurity strength is tuned by shifting the on-site energy $U_{ss}$ and $U_{\tilde p \tilde p}$ of the corresponding orbital to lower values (horizontal axis).
    (a) Full spectrum. 
    Shaded regions indicate the three bulk bands.
    The in-gap states are not shown for better readability.
    (b) Zoom of the gap between the first and second band. 
    The blue circles label resonances corresponding to ring states which remain pinned inside the gap, while in-gap resonances with $\tilde{p}$ weight (orange circles) gradually hybridize with and enter the lower band as the impurity strength increases.
    }
    \label{fig:ring_states_lattice}
\end{figure}

In Fig.~\ref{fig:ring_states_lattice} we show the impurity-induced state frequencies for a finite lattice of size $21 \times 21$ for our three-band tight-binding model.
% Here, we focus on states induced by $s$ and $p$
We add one rank-2 impurity at the center of the lattice by shifting the onsite-energy of both $s$ and $\tilde p$ orbital on that site to lower values.
Here, we choose $U_{ss}=U_{\tilde p \tilde p}$.
Fig.~\ref{fig:ring_states_lattice}(a) displays the full spectrum.
The local impurity induces states below the lowest band, which are highly localized at the impurity site and rapidly decrease in frequency as the impurity strength is increased.
The existence of these states is always guaranteed independent of the existence of ring states in gaps.
% The fact that the $\tilde{p}$ impurity states
We observe that, consistent with the in-gap zero-crossing of the $s$-dominated eigenvalue of $g$ [Fig.~\ref{fig:SI_greens_function_TB_3band}(b)], the impurity generates in-gap states in the gap between the first and second band whose energy saturates as the impurity strength is increased [Fig.~\ref{fig:ring_states_lattice}(b)].
The mirror even--odd band inversion forces an impurity affecting the $s$ orbital to host robust in-gap ring states.
In contrast, the $\tilde{p}$ impurity-induced in-gap states are not pinned and merge into the first band for a sufficiently strong impurity.
Impurities affecting the $\tilde{p}$ orbital generate only accidental in-gap modes which is in line with the fact that there is no in-gap zero crossing of the $\tilde p$-dominated eigenvalues of $g$ [Fig.~\ref{fig:SI_greens_function_TB_3band}(b)].

\section{Adiabatically transforming the three-band model into an effective two-band model}
\label{sec:adiabatic_transformations}

\begin{figure}
    \centering
    \includegraphics{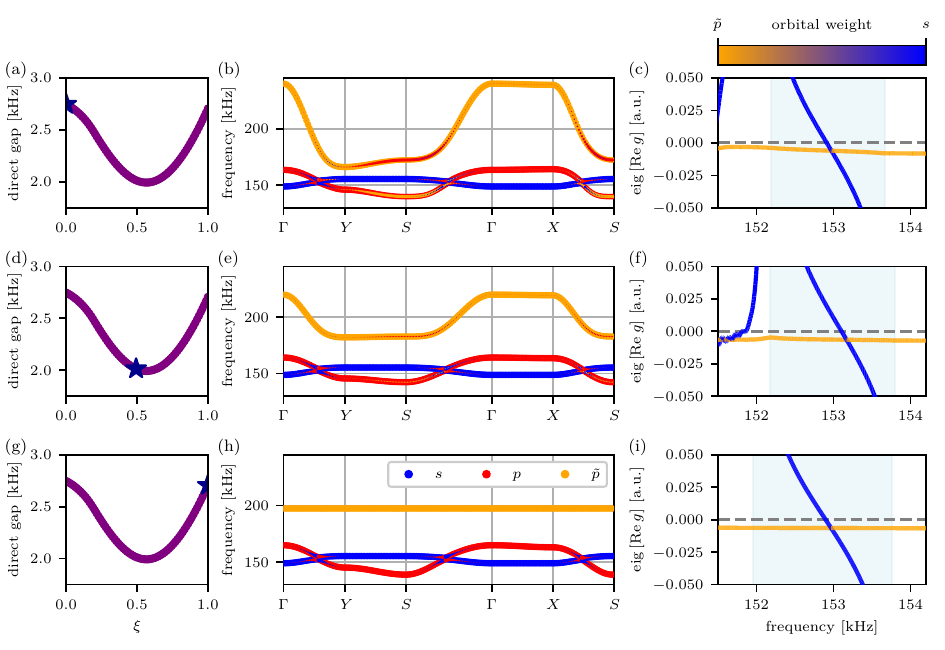}
    \caption{{Adiabatic interpolation from the three-band model to an effective $s$/$p$ two-band model and evolution of the impurity-projected Green's function eigenvalues.} 
    (a), (d), (g) Direct band gap between the lower two bands as a function of the interpolation parameter $\xi$, demonstrating that the gap remains finite along the symmetry-preserving interpolation defined in Eq.~(\ref{eq:2band_3band_interpolation}).
    % The light blue star labels the interpolation parameter $\xi = 0$.
    (b), (e), (h) Band structure along a high-symmetry path in the Brillouin zone for three representative points of the interpolation: 
    (b) $\xi=0$ (star in (a); three-band model), (e) $\xi=0.5$ (star in (d)), and $\xi=1$ (star in (g); effective two-band limit with a decoupled $\tilde p$ band).
    The blue, red and orange markers indicate the orbital character $s$, $p$ and $\tilde p$, respectively.
    (c), (f), (i) Corresponding eigenvalues of the real parts of the orbital-projected Green's function $g$ evaluated using Eq.~(\ref{eq:imp_projected_G_numerics_rank2_numerics}).
    Light blue shading marks the full band gap between the first and the second band and the dashed line indicates ${\rm eig}[{\rm Re}\,g]=0$.
    The colors indicate the orbital weight computed as the overlap of the corresponding eigenvector $\ket{v_{{\rm real}}}$ of ${\rm Re} \, g$ with the $s$ orbital $|\langle s | v_{{\rm real}} \rangle|^2$.
    Blue corresponds to $|\langle s | v_{{\rm real}} \rangle|^2=1$, whereas orange corresponds to $|\langle s | v_{{\rm real}} \rangle|^2=0$ and therefore maximal overlap with the $\tilde p$ orbital.
    }
    \label{fig:FigSI_adiabatic_path}
\end{figure}

The dynamical matrix of the three-band model can be smoothly transformed to an effective $s$-$p$ two-band dynamical matrix, with an entirely decoupled $\tilde{p}$ band and without closing the band gap of the two lower bands.
The symmetry-preserving transformation smoothly exchanges the three-band and two-band tight-binding parameters $t_{\alpha \beta}^{ij}$ listed in Tab.~\ref{app:hoppings} using the following linear interpolation
\begin{equation}
\begin{aligned}
    t_{\alpha \beta}^{ij} &\mapsto t_{\alpha \beta, 3}^{ij} (1-\xi) + t_{\alpha \beta, 2}^{ij} \tau \\
    \omega_\alpha &\mapsto \omega_{\alpha, 3} 
\end{aligned}
\label{eq:2band_3band_interpolation}
\end{equation}
where  $\alpha, \beta \in \{ s, p, \tilde{p}\}$, $i \in \{ 0, 1, 2\}$, $j \in \{ 0, 1\}$ and the subscripts $2$ and $3$ indicate parameters corresponding to the the two- and three-band model, respectively. 
$\xi \in [0, 1]$ is the interpolation parameter and $\tau=\xi$ if a parameter $t_{\alpha \beta, 3}^{ij}$ has a corresponding $t_{\alpha \beta, 2}^{ij}$ parameter (see Tab.~\ref{app:hoppings}), else $\tau=0$.
The evolution of the bulk spectrum along the path defined in Eq.~(\ref{eq:2band_3band_interpolation}) is illustrated in Fig.~\ref{fig:FigSI_adiabatic_path}:
the direct gap between the first and the second band remains finite for all $\xi$, and representative band structures for $\xi=0$, $\xi=0.5$ and $\xi=1$ are shown in Fig.~\ref{fig:FigSI_adiabatic_path}(b), (e), (h).
Throughout the interpolation, the orbital character of the two lower bands remains predominantly $s$/$p$, while the $\tilde p$ weight is shifted into the high-frequency band.
For the same representative values of $\xi$, Fig.~\ref{fig:FigSI_adiabatic_path}(c), (f), (i) show the corresponding eigenvalues of the real part of the impurity-projected Green's function $g$ across the full band gap between the first and the second band.
The zero-crossing of the $s$-dominated eigenvalues in the lower gap persists along the adiabatic path, whereas the $\tilde p$-dominated eigenvalues remain negative in the gap.

\section{Adiabatic swap within the mirror-odd subspace}
\label{sec:hot_swap}

\begin{figure}
    \centering
    \includegraphics{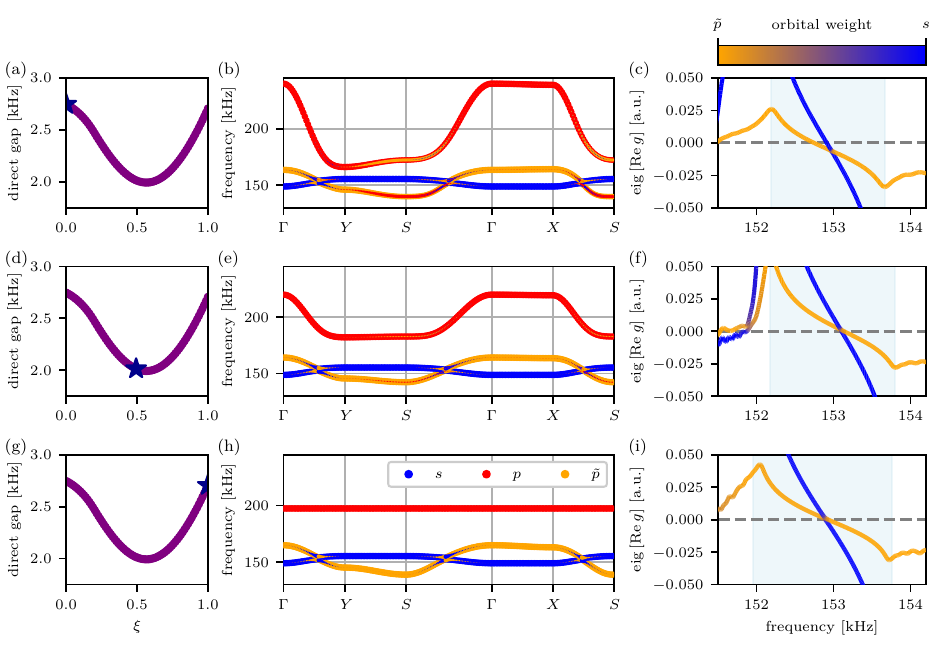}
    \caption{{Adiabatic interpolation from the swapped three-band model to an effective $s$/$\tilde p$ two-band model and evolution of the impurity-projected Green's function eigenvalues.}
    (a), (d), (g) Direct band gap between the lower two bands as a function of the interpolation parameter $\xi$, demonstrating that the gap remains finite along the symmetry-preserving interpolation defined in Eq.~(\ref{eq:2band_3band_interpolation}).
    % The light blue star labels the interpolation parameter $\xi = 0$.
    % Stars mark the three representative values of $\xi$ used in (b), (c), (e), (f) and (h), (i).
    (b), (e), (h) Band structure along a high-symmetry path in the Brillouin zone for representative interpolation points: 
    (b) $\xi=0$ (star in (a); three-band model corresponding to the dynamical matrix in Eq.~(\ref{eq:dynamical_matrix_rotated}), obtained after the adiabatic swap within the mirror-odd subspace), (e) $\xi=0.5$ (star in (d)), and $\xi=1$ (star in (g); effective two-band limit with a decoupled $p$ band).
    The blue, red and orange markers indicate the orbital character $s$, $p$ and $\tilde p$, respectively.
    (c), (f), (i) Corresponding eigenvalues of the real parts of the orbital-projected Green's function $g$ evaluated using Eq.~(\ref{eq:imp_projected_G_numerics_rank2_numerics}).
    Light blue shading marks the full band gap between the first and the second band and the dashed line indicates ${\rm eig}[{\rm Re}\,g]=0$.
    The colors indicate the orbital weight computed as the overlap of the corresponding eigenvector $\ket{v_{{\rm real}}}$ of ${\rm Re} \, g$ with the $s$ orbital $|\langle s | v_{{\rm real}} \rangle|^2$.
    Blue corresponds to $|\langle s | v_{{\rm real}} \rangle|^2=1$, whereas orange corresponds to $|\langle s | v_{{\rm real}} \rangle|^2=0$ and therefore maximal overlap with the $\tilde p$ orbital.}
    \label{fig:FigSI_adiabatic_path_swap}
\end{figure}

We can exchange the two mirror-odd orbitals $p$ and $\tilde p$ by smoothly transforming the dynamical matrix into an effective $s$-$\tilde p$ two-band dynamical matrix, with an entirely decoupled $p$ band and without closing the band gap of the two lower bands.
The key idea is to use a unitary transformation of the three-band dynamical matrix in the fixed $\{s, p, \tilde{p}\}$ basis.
The unitary transformation 
\begin{align*}
        \tilde{\mathcal D}_\mu(\vec k) = U_\mu \mathcal D(\vec k) U_\mu^\dagger
\end{align*}
with
\begin{align}
    U_\mu = \begin{bmatrix}
    1 & 0 & 0 \\
    0 & \cos{\theta(\mu)} & \sin{\theta(\mu)} \\
    0 & -e^{{\rm i} \pi \mu} \sin{\theta(\mu)} & e^{{\rm i} \pi \mu} \cos{\theta(\mu)}
    \end{bmatrix}, ~ \theta(\mu) = \frac{\pi}{2} \mu, ~ \mu \in [0, 1]
    \label{eq:unitary}
\end{align}
continuously transforms the $p$ into the $\tilde{p}$ matrix elements and vice versa, leaving the $s$ orbital matrix elements unchanged.
At the same time, since it is a unitary transformation, the eigenvalues of $\mathcal{D}(\vec k)$ are preserved.
At $\mu=1$, where
\begin{equation}
    U_{\mu=1} = \begin{bmatrix}
    1 & 0 & 0\\
    0 & 0 & 1\\
    0 & 1 & 0
    \end{bmatrix},
\end{equation}
we arrive at the dynamical matrix
\begin{equation}
    \tilde{\mathcal{D}}_{\mu=1}(\vec k) = \begin{bmatrix}
    \mathcal s({\vec k}) & \mathcal f_{s\tilde p}({\vec k})& \mathcal f_{sp}(\vec k)\\
    \mathcal f_{s\tilde p}^{*}({\vec k}) & \mathcal {\tilde p}({\vec k})& \mathcal f_{p \tilde p}^{*}(\vec k)\\
    \mathcal f^*_{sp}({\vec k}) & \mathcal f_{p\tilde p}({\vec k})& \mathcal {p}(\vec k) 
    \end{bmatrix}.
    \label{eq:dynamical_matrix_rotated}
\end{equation}
Note that this dynamical matrix corresponds to an exact exchange of $p$ and $\tilde p$ matrix elements only if $f_{p\tilde p}^{*}({\vec k})=f_{p\tilde p}({\vec k})$, i.e., $f_{p\tilde p}({\vec k})$ is real.
In our case $f_{p\tilde p}^{*}(k_x, k_y)=f_{p\tilde p}(k_x, -k_y) \neq f_{p\tilde p}(k_x, k_y)$.
We then apply the transformation (\ref{eq:2band_3band_interpolation}) to the dynamical matrix in Eq. (\ref{eq:dynamical_matrix_rotated}), such that $f_{p\tilde p}({\vec k})$ and $f_{p\tilde p}^{*}({\vec k})$ are turned off adiabatically, yielding an effective $s$-$\tilde p$ two-band model with an entirely decoupled third $p$ band.

The gapped interpolation to an effective $s$/$\tilde p$ two-band model by decoupling the $p$ band after the adiabatic swap within the mirror-odd subspace is illustrated in Fig.~\ref{fig:FigSI_adiabatic_path_swap}.
The zero-crossing of the $s$-dominated eigenvalues in the lower gap persists both along the adiabatic swap and the transformation that decouples the $p$ band [Fig.~\ref{fig:FigSI_adiabatic_path_swap}(c), (f), (i)].
After the swap, the $\tilde p$ orbital participates in the band inversion of the two lower bands [Fig.~\ref{fig:FigSI_adiabatic_path_swap}(b), (e), (h)].
Hence, in addition to the zero-crossing of the $s$-dominated eigenvalues in the lower gap, we obtain a zero-crossing of the $\tilde p$-dominated eigenvalues [Fig.~\ref{fig:FigSI_adiabatic_path_swap}(c), (f), (i)].

%TC:endignore

\end{document}